\documentclass{article}

\usepackage{arxiv}
\usepackage[T1]{fontenc}    

\usepackage{url}            
\usepackage{booktabs}       
\usepackage{amsfonts}       
\usepackage{nicefrac}       
\usepackage{microtype}      
\usepackage{doi}

\usepackage[utf8]{inputenc}
\usepackage{graphicx}
\usepackage{float}
\usepackage{caption}
\usepackage{subcaption}
\usepackage[pdftex]{color}
\usepackage{mwe}
 \usepackage{multirow}
\usepackage{hyperref}
\usepackage{amsmath}
\usepackage{pifont}
\usepackage{amssymb}
\usepackage[dvipsnames]{xcolor}
\usepackage[ruled]{algorithm2e}
\usepackage{tcolorbox}
\usepackage{lipsum}
\tcbuselibrary{skins,breakable}
\usetikzlibrary{shadings,shadows}
\usepackage{xcolor}
\usepackage{cancel}
\usepackage{wrapfig,lipsum,booktabs}
\usepackage{natbib}

\newcounter{example}[section]

\newtheorem{definition}{Definition}
\newtheorem{remark}{Remark}

\title{Clustered Mallows Model}

\author{ \href{https://orcid.org/0000-0002-3487-2961}{\includegraphics[scale=0.06]{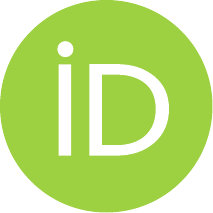}\hspace{1mm}Luiza S.C. Piancastelli} \\
	School of Mathematics and Statistics\\
	University College Dublin\\
	Dublin, Ireland \\
	\texttt{luiza.piancastelli@ucd.ie} \\
	\And
	\href{https://orcid.org/0000-0003-4778-0254}{\includegraphics[scale=0.06]{orcid.pdf}\hspace{1mm} Nial Friel} \\
    School of Mathematics and Statistics\\
    Insight Centre for Data Analytics \\
	University College Dublin\\
	Dublin, Ireland \\
	\texttt{nial.friel@ucd.ie} \\
}

\date{}

\renewcommand{\headeright}{ }
\renewcommand{\shorttitle}{CMM}

\hypersetup{
pdftitle={Clustered Mallows Model},
pdfsubject={q-bio.NC, q-bio.QM},
pdfauthor={Luiza S.C. Piancastelli, Nial Friel},
pdfkeywords={Mallows model, ranking data, Bayesian learning},
}

\begin{document}
\maketitle

\begin{abstract}
	Rankings are a type of preference elicitation that arise in experiments where assessors arrange items, for example, in decreasing order of utility. Orderings of $n$ items labelled $\{1, \ldots, n\}$ denoted by $\pi$ are permutations ($\pi \in \mathcal{P}_n$) that reflect strict preferences. That is, an item ranked $i^{th}$ is more preferred to the judge than any other in position $j$ if $j >i$. For a number of reasons, strict preference relations can be unrealistic assumptions for real data. 
For example, when items share common traits it may be reasonable to attribute equal ranks among these items. 
Depending on the situation, there can be different importance attributions to decisions that form $\pi$. In a situation with, for example, a large number of items, an assessor may wish to rank at top a certain number items; to rank other items at the bottom and to express indifference to all other items. 
In this type of \textit{top/bottom} elicitation, middle-rank alternatives can be a close to uniform placement of the remaining items. In addition, when aggregating opinions, a judging body might be decisive about some parts of the rank but ambiguous for others.
In this paper we extend the well-known Mallows \citep{Mallows} model (MM) to accommodate item indifference, a phenomenon that can be in place for a variety of reasons, such as those above mentioned.
The underlying grouping of similar items motivates the proposed Clustered Mallows Model (CMM). The CMM can be interpreted as a Mallows distribution for tied ranks where ties are learned from the data. Although the idea of tied Mallows models dates back to \cite{chung_marden}, it has not been developed other than in the hypothesis testing context. The CMM provides the flexibility to combine strict and indifferent relations, achieving a simpler and robust representation of rank collections in the form of ordered clusters. Bayesian inference for the CMM is in the class of doubly-intractable problems since the model's normalisation constant is not available in closed form. We overcome this challenge by sampling from the posterior with a version of the exchange algorithm \citep{murray2006}. A pseudo-likelihood approximation is also provided as a computationally cheaper alternative. Information-criterion and data-based selection strategies are outlined to select between partitions of different sizes. Real data analysis of food preferences and results of Formula 1 races are presented, illustrating the CMM in practical situations.
\end{abstract}


\section{Introduction}\label{sec:intro}

The exercise of ranking a list of items, for example, from best to worst, is a common situation in a variety of contexts. 
Item ranking are commonly employed in market research, for example. While many other real-life situations  naturally produce rankings. One such example is the Single Transferable Voting system (STV) adopted in countries such as Ireland, Australia and the United Kingdom (local elections in Scotland and Northern Ireland). In this form of election, voter preferences are given in the form of a ranked-choice ballot. In sports, they represent the result of racing competitions or competitors' places in long-term championships. Rankings also underlie many recommendation systems, such as the order of pages in web searches.

Mathematically, a ranking of $n$ alternatives labelled $\{1, \dots, n\}$ is an observation in $\mathcal{P}_n$, the space of permutation of $n$ labels. This will be denoted by $\pi$. Important operations on $\pi$ are $\pi(i)$ and $\pi^{-1}(i)$. The first corresponds to the item label ranked $i^{th}$ and its inverse, $\pi^{-1}(i)$, is the rank of the object $i$. For example, if $\pi = (3,1,2)$, then $\pi(1) = 3$ means that the item 3 ranked first and $\pi^{-1}(1)=2$ means that item 1 is ranked second.

Statistical and machine learning approaches for ranking data go beyond complete rankings, that is, $\pi \in \mathcal{P}_n$. For instance, pairwise preference studies gather opinions by asking the assessor to choose between item pairs. This approach elicits preferences for pairs $\{i, j\}, i\neq j$ taken from $\{1, \ldots, n\}$, denoted as $i \succ j$ or $j \succ i$. The symbol $\succ$ denotes a strict preference for the first over the second item and multiple comparisons are often presented to the same assessor. For example, suppose that three pairwise comparisons of $\{1,\ldots,4\}$ are given as $\mathcal{C} = \{ 4\succ 2, 1 \succ 3, 2 \succ 1\}$, then $\mathcal{C}$ implies that $\pi = (4,2,1,3)$. Most often, $\mathcal{C}$ does not contain all comparisons needed to explicit a complete ranking. This happens necessarily when $\# \mathcal{C} < (n-1)$, where $\# \mathcal{C} $ is the cardinality of $\mathcal{C}$. In this case, more than one $\pi$ will be compatible with $\mathcal{C}$ if opinions are consistent. For example, $\mathcal{C} = \{2 \succ 1, 3 \succ 4\}$ induces $4!/(2! 2!) = 6$ complete permutations including $\pi_1 = (2,1,3,4)$, $\pi_2 = (3,4,2,1)$, $\pi_3 = (3,2,4,1)$, each of which are consistent with $\mathcal{C}$. Pairwise preference sets are one possible generating mechanism of incomplete ranking data. The partial observation of $\pi$ will be denoted by $\widetilde{\pi}$ to indicate that the complete ordering is not observed. Another type of partial observation $\widetilde{\pi}$ of $\pi$ arises in the context of, what we term top$-k$ elicitation. In this setting, assessors provide their first $k<n$ choices, determining $\pi(1), \ldots, \pi(k)$. The single transferable voting is an example of this situation. See also \cite{gormley2006} for another example in the context of Irish third-level college applications. 

 
Methodology to handle $\pi$ and $\widetilde{\pi}$ is gaining recent attention in Statistics and Machine Learning (ML). In the latter, this is known as \textit{learning to rank approaches} and some popular approaches are \cite{burges2006learning}, \cite{burges2010ranknet}. In recommendation systems, ranking-based methods are often advocated due to user inconsistency or limited scale of ratings (\cite{brun2010}, \cite{brun2011}, \cite{lu_boutilier2014}). 

In this paper we take a statistical approach to model $\pi$ (or $\widetilde{\pi}$) by describing it in terms of a probability distribution. Probabilistic models describe the underlying mechanism that generates $\pi$ by making different assumptions about the assessor's decision process. \cite{prob_ranks} categorizes these  as (1) Thurstone order models, (2) pairwise ranking models, (3) distance-based, and (4) multistage ranking models. The latter is a popular approach where $\pi$ is formed from sequential decisions on decreasing item sets. Multistage ranking is underpinned by the \textit{independence of irrelevant alternatives (IIA)} assumption. Under IIA, the probability of selecting an object from a set is not affected by the presence or absence of others. One of the most famous models in this class is the Plackett-Luce \citep{plackett1975}.

Our work relates to category (3), for which the Mallows' model \citep{Mallows} is the best known example. Distance-based models attribute probability to $\pi$ according to its distance to a population consensus $\pi_0$. This is written as $d(\pi, \pi_0)$ where $d(\cdot, \cdot)$ is a distance in $\mathcal{P}_n$, and $\pi_0$ is an unknown model parameter. The Mallows model (MM) is given by
\begin{equation}\label{eq:mallows_model}
   p(\pi|\pi_0, \alpha) = \frac{\exp\{- \alpha d(\pi, \pi_0) \} }{\phi(\alpha,  \pi_0)}, 
\end{equation}
where $\alpha >0$ governs the spread around $\pi_0 \in \mathcal{P}_n$. It is an exponential family model where $\pi_0$ and $\alpha$ are estimated from the data and $\phi(\alpha,  \pi_0)$ is the normalisation constant. In the Mallows$-\phi$ model, $d(\cdot, \cdot)$ is the Kendall's $\tau$ distance in $\mathcal{P}_n$, which counts the number of pairwise disagreements in $\pi$ and $\pi_0$. Other popular distance choices are the Hamming and Cayley distance metrics. The term $\phi(\alpha, \pi_0)$ is often written as $\phi(\alpha)$ because common 
choices for $d$ are unaffected by permutations of item labels (i.e., are right-invariant.)

The majority of applications and development of (\ref{eq:mallows_model}) uses the Kendall, Hamming, or Cayley forms of $d(\cdot, \cdot)$. This is because the normalisation term $\phi(\alpha,  \pi_0)$ has a simple analytical form in these cases. Expressions for $\phi(\alpha)$ were derived by \cite{fligner1986} under the Kendall and Hamming distances, and later for the Cayley distance by \cite{irurozki2018sampling}. These are each non-trivial advancements that do not generalise to other distances. 

As a consequence, other distances such as the Spearman and Footrule make (\ref{eq:mallows_model}) difficult to evaluate. Computing $\phi(\alpha,  \pi_0) = \sum_{\pi \in \mathcal{P}_n} \exp\{ - \alpha d(\pi, \boldsymbol{\pi}_0) \}$ exactly requires enumerating every $\pi \in \mathcal{P}_n$. This is unfeasible except for trivially small $n$, rendering the MM an intractable probability distribution in most cases.

One important extension of (\ref{eq:mallows_model}) is the Generalised Mallows Model (GMM, \cite{fligner1986}).
The GMM models the importance of different ranking stages with $\boldsymbol{\alpha} = (\alpha_1, \ldots, \alpha_{n-1})$ according to $p(\pi|\boldsymbol{\alpha}) \propto \exp \left( - \sum_{j=1}^{n-1} \alpha_j d^j(\pi, \pi_0)\right)$ relying on the decomposition $d(\pi, \pi_0) = \sum_{j=1}^{n-1} d^j(\pi, \pi_0)$. The latter assumes that the total distance can be expressed as the sum of independent comparisons between $\pi$ and $\pi_0$ that occur at ranking stages $j = 1, \ldots, n-1$, i.e., $d(\cdot,\cdot)$ is \textit{decomposable}. In another direction, \cite{meila2008}, \cite{meila2010} introduced the Infinite Generalised Mallows model (IGM) for top$-k$ for partial permutations $\widetilde{\pi}$. More details about the IGM and its connection to our approach are explored later. In \cite{crispino2017}, the MM is extended to $\widetilde{\pi}$ from pairwise comparisons with inconsistencies. Recent work by \cite{vitelli2018} approaches modeling a collection of independent ranks $(\pi_1, \ldots, \pi_q)$ from $q$ non-homogeneous assessors. This is denoted by $\underline{\pi}$, a $(q \times n)$ matrix. The authors propose modelling $\underline{\pi}$ with a finite mixture of Mallows' distributions to find cluster-specific consensus and spreads. 

The study of $\underline{\pi}$ is a topic of emerging interest, often referred to as \textit{rank aggregation}. See \cite{deng2014}, \cite{zhu2023}, \cite{li2020extended}, for example. The contribution of this work goes in this direction. We propose an extension of the MM for situations where $\underline{\pi}$ does not convey a strict preferences consensus. The parameter of main interest in the Mallows model is $\pi_0 \in \mathcal{P}_n$, a complete and strict ordering of $\{1, \ldots, n\}$. Estimating $\pi_0$ means inferring $i \succ j$ or $j \succ i$ for all $i,j$ which could be unrealistic. The aim of the model proposed in this paper is to accommodate indifference between some items $i,j$, denoted by $i \sim j$. This can come from the data's underlying generating process such as top-bottom elicitation or convey uncertainty between assessors. This is made clear in the following subsection. 

\subsection{A motivating example}

\begin{figure}
    \centering
    \includegraphics[width = 0.65\linewidth]{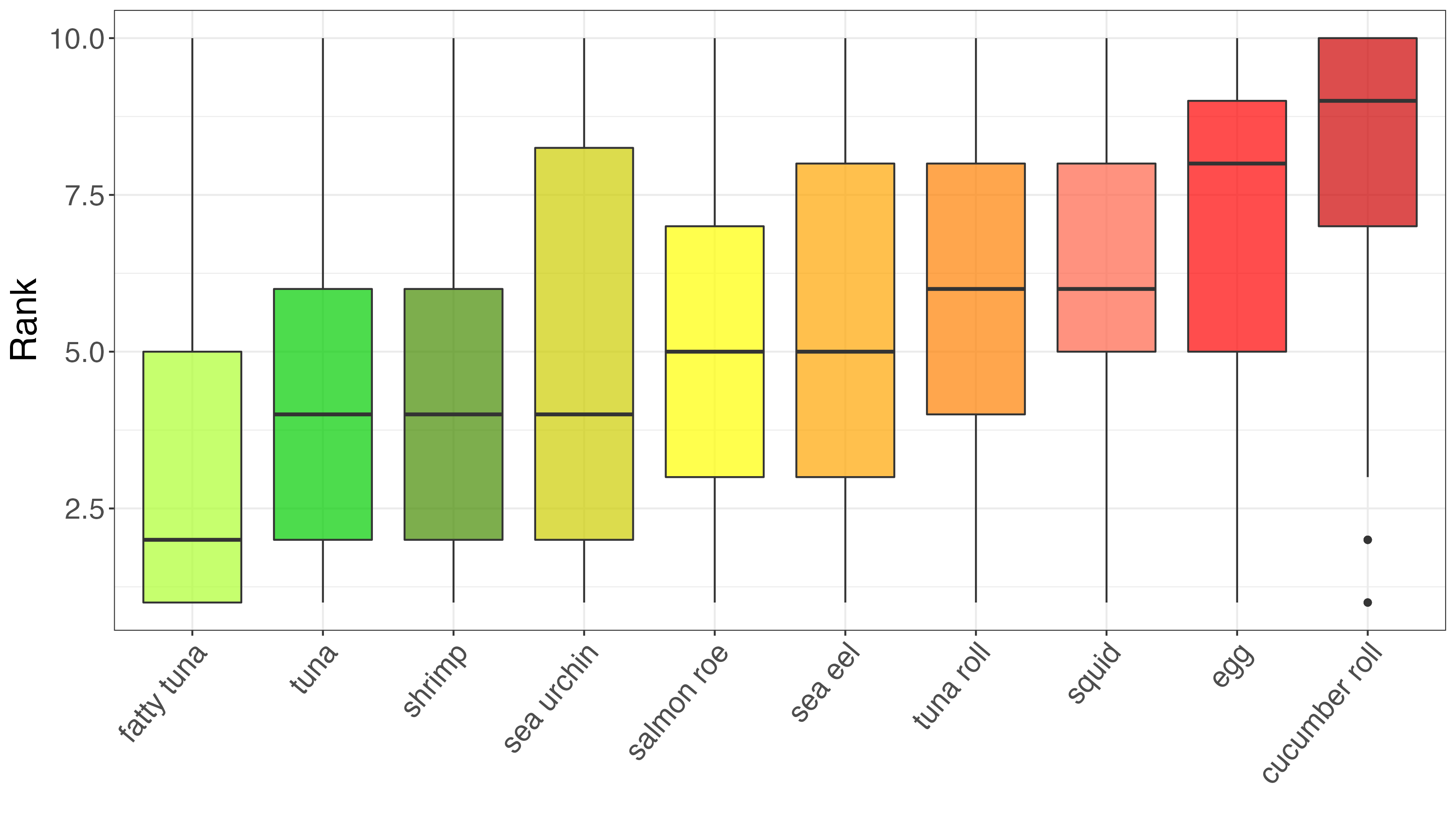}
    \caption{Distribution of ranks given to sushi variants by respondents who lived in Japan's region Tohoku at least until 15 years old.}
    \label{fig:tohoku}
\end{figure}

\cite{sushi} surveys food preference in Japan by asking participants to rank 10 varieties of sushi. This is a famous ranking data set, also analysed in \cite{vitelli2018}. According to \cite{sushi}, food taste in Japanese culture is largely influenced by geographical regions. Accordingly, participants are separated by their response to the question \textit{"What is the region in which you have been the most longly lived until 15 years old?".} The subset of rankings from \texttt{Tohoku} contains $n=280$ observations and is illustrated in Figure \ref{fig:tohoku}. Boxes showcase the marginal distribution of rankings given to the types of sushi, obtained fixing $i \in \{$\texttt{fatty tuna}, \ldots, \texttt{cucumber roll}\} and gathering $\pi_1^{-1}(i), \ldots, \pi_{280}^{-1}(i)$. Arranging the boxplots in increasing order of median value gives a type of summary of the opinions within $\underline{\pi}$. 

\begin{figure}
\begin{minipage}{0.5\textwidth}
\includegraphics[width=1.0\linewidth,keepaspectratio=true]{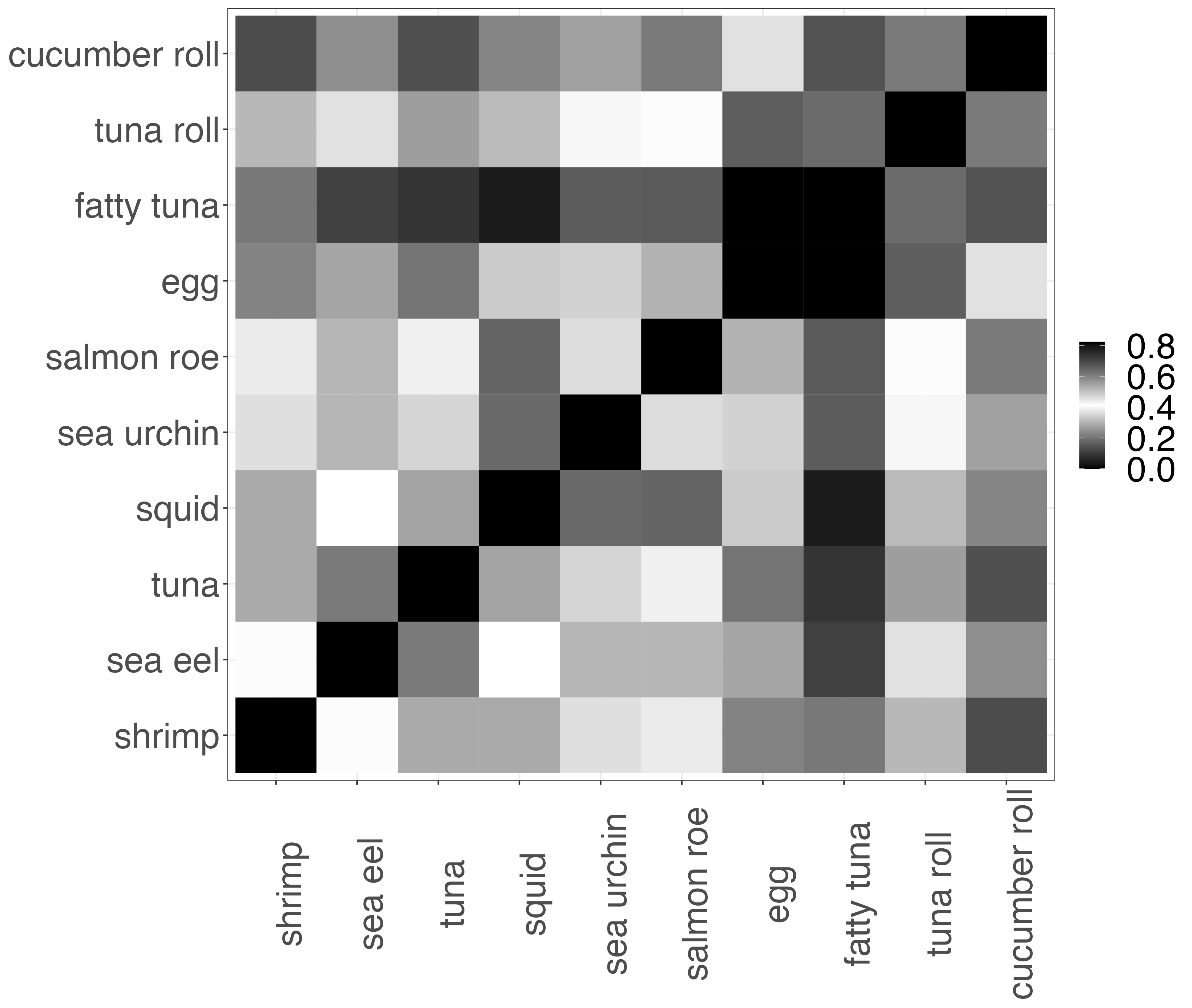}

\end{minipage}
\hfill
\begin{minipage}{0.45\textwidth}
\scriptsize
\centering
\begin{equation*}
\underline{\pi} =
\begin{bmatrix}
3 & 1 & 7 & 9 & 8 & 10 & 2 & 6 & 5 & 4 \\
9 & 1 & 4 & 5 & 6 & 2 & 3 & 8 & 10 & 7\\
1 & 5 & 9 & 2 & 8 & 3 & 4 & 10 & 6 & 7 \\
\vdots & \vdots & \vdots & \vdots& \vdots& \vdots& \vdots& \vdots& \vdots& \vdots\\
8 & 3 & 1 & 2 & 9 & 4 & 7 & 10 & 6 & 5\\
3 & 9 & 4 & 1 & 6 & 2 & 7 & 8 & 10 & 5 \\
1 & 6 & 4 & 5 & 8 & 3 & 7 & 10 & 9 & 2 \\
\end{bmatrix}
\end{equation*}
\tiny
\begin{equation*}
\begin{bmatrix}
\texttt{tuna} & \texttt{shrimp} & \cdots  &  \texttt{sea urchin} &  \texttt{squid} \\
\texttt{tuna roll} & \texttt{shrimp} & \cdots  &  \texttt{cucumber roll} & \texttt{egg} \\
\texttt{shrimp} & \texttt{sea urchin} & \cdots & \texttt{salmon roe} &  \texttt{egg} \\
\vdots  & \vdots& \vdots& \vdots& \vdots\\
\texttt{fatty tuna} & \texttt{tuna} & \cdots  &  \texttt{salmon roe} & \texttt{sea urchin} \\
\texttt{tuna} & \texttt{tuna roll} & \cdots  &  \texttt{cucumber roll} & \texttt{sea urchin} \\
\texttt{shrimp} & \texttt{salmon roe} & \cdots & \texttt{tuna roll} &  \texttt{sea eel} \\
\end{bmatrix}
\end{equation*}

\null
\par\xdef\tpd{\the\prevdepth}
\end{minipage}

\caption{On the right, the proportion of times that $i$ (rows) is ranked below $i'$ (columns) in $\underline{\pi}$. To the left, the top and bottom rows of $\underline{\pi}$ are shown. The top matrix takes items labeled as $\{1, \ldots, 10\}$ which corresponds to \texttt{shrimp}, \texttt{sea eel}, \texttt{tuna}, \texttt{squid}, \texttt{sea urchin}, \texttt{salmon roe}, \texttt{egg}, \texttt{fatty tuna}, \texttt{tuna roll} and \texttt{cucumber roll} as exemplified at the bottom.}\label{fig:sushi_matrix}
\end{figure}

Figure \ref{fig:tohoku} is a much simplified representation of the data however. 
We highlight that the structure of $\underline{\pi} = (\pi_1, \ldots, \pi_{280})$ is in fact much more complex since $\pi_j$ is multivariate. Naturally, elements of a ranking carry relative information about the item set and this comes from imposing $\pi_j \in \mathcal{P}_n$. Figure \ref{fig:sushi_matrix} provides an alternative visualisation of $\underline{\pi}$ that brings attention to this fact. It is constructed by gathering how many times sushi $i$ is ranked lower that sushi $i'$ for all $\pi_j$ and constructing the empirical proportion, $\bar{p}(i, i') =  \sum_{j=1}^{280} I\{ \pi_j(i) \succ \pi_j(i') \}/280$. In Figure \ref{fig:sushi_matrix}, $i$ is shown as rows and $i'$ as columns. 
To the right, a snapshot of the data is presented where the different sushi variants $\texttt{shrimp,sea eel,tuna,squid,sea urchin,salmon roe,egg,fatty tuna,tuna roll,} $\newline $\texttt{cucumber roll}$ are labeled $\{1, \ldots, 10\}$.

Figure \ref{fig:tohoku} immediately suggests \texttt{fatty tuna} and \texttt{cucumber roll} as the best and worst options. However, the overall opinion regarding middle ranks is less clear. This is supported by the preliminary analysis shown in Figure \ref{fig:sushi_matrix}, where the proportion is close to 0.5 for the following pairs: \texttt{\{tuna roll, squid\} $\approx 0.51$}, \texttt{\{sea urchin, sea eel\}} $\approx 0.52$, \texttt{\{sea eel, salmon roe\}} $\approx 0.52$ and \texttt{\{tuna, shrimp\}} $\approx 0.54$. In these cases, empirically, at least, there appears to be an indifference in the preference of one over the other. 
The aim of our statistical model is accommodate situations like this, where there may be ties (or indifference) in certain items. 



\subsection{Modelling indifference}

The proposed model is an extension of the MM given in equation (\ref{eq:mallows_model}) that allows the consensus $\pi_0$ to contain ties. As before we denote tied items $i$ and $j$ by $i \sim j$ to indicate that there is not a strict preference between them. This can be interpreted as replacing $\pi_0$ with an ordered partition of the item set $\{1, \ldots, n\}$ where alternatives in the same group are exchangeable. This motivates naming our approach the Clustered Mallows Model (CMM). Within the CMM framework, clustering occurs on the item set, contrasting with \cite{vitelli2018}, where assessors are grouped instead.

To achieve this, each item is associated with one of $L \leq n$ clusters or groups. We begin by assuming that the number of ordered clusters is fixed. The choice of $L$ will be addressed later on. We introduce the notation  $\boldsymbol{z} = (z(1), \ldots, z(n))$, a vector of unknown cluster allocation of $i \in \{1, \ldots, n\}$, where $z(i)$ denotes the ordered cluster allocation of item $i$ where $z(i) \in \{1, \ldots, L\}$. The inverse operation $z^{-1}(l)$ means listing objects that were assigned label $l$, defined for $l \in \{1, \ldots, L\}$. A rank interpretation of cluster labels is fixed by imposing the following two conditions: 
\begin{description}
    \item[Condition (I):] $i \succ j$ if $z(i)\succ z(j)$.
    \item[Condition (II):] $i \sim j$ if $z(i) = z(j)$. 
\end{description}
The first condition is interpreted as \textit{item $i$ is preferred over any other object with a higher cluster label}. While the second imposes indifference between items within the same cluster, so that if $i \sim j$ then $z(i)$ and $z(j)$ are equal. 

As an example, take $\boldsymbol{z} = (2,1,1,2,3)$. This allocation implies that items 2 and 3 are in the first cluster since $z(2) = z(3) =1$. They are also both top preferences since $2 \sim 3$ and are preferred over the remaining items. The clusters $( \{z^{-1}(1)\}, \{z^{-1}(2)\}, \{z^{-1}(3)\} )$ are $\{2,3\}, \{1,4\}, \{5\}$. A useful notation for clustered groups is the \textit{Brack operation} \citep{marden1995}. Denoted here by $\{\boldsymbol{z}\} = ( \{z^{-1}(1)\}, \ldots, \{z^{-1}(L)\} )$, clusters are placed in increasing order of cluster label and $\{ \cdot \}$ indicates interchangeability of items within a cluster. Hence, $\{ \boldsymbol{z} \} = (\{2,3\}, \{1,4\}, \{5\})$ in this example.

Modelling $\underline{\pi}$ in terms of $\boldsymbol{z}$ can be advantageous over the MM 
when preferences are not all necessarily strict. A straightforward example is when alternatives share common features or simply when assessors are jointly indecisive about ranking some items. In such cases, relative preferences $\bar{p}(i, i')$ will be close to 0.5 as in our motivating example. In addition, it is reasonable to assume some interchangeability when $n$ is moderate or large. 

Inferring $\boldsymbol{z}$ from $\underline{\pi}$ with the CMM provides answers to the following. Do evaluations determine strict preferences between all $i,j \in \{1, \ldots, n\}$? If yes, then $L =n$, and the CMM has the MM as a special case. Otherwise, the immediate next question is  how many clusters are there and to which group does each object belong to? 

Learning $\boldsymbol{z}$ is the process of determining $L$ and $n_1, \ldots, n_L$, where $n_l = \sum_{i=1}^n I\{ z(i)=l\}$ is the size of cluster $l$. If $n_1, \ldots, n_L$ is free to vary, then indifference is accommodated in any part of the rankings. One setting that can result from this kind of flexibility in this \textit{'best-worse scaling'}. This is a choice mechanism that recognizes extremes to be more easily recognized by individuals (\cite{doignon2023}, \cite{marley2016}, \cite{MARLEY201224}). A CMM with $L$ clusters and $n_l = 1$ for small and large $L$ reflects best-worse scaling, for example, $n_1 =1, n_2=8, n_3 =1$ encodes the top and the bottom choices of $n=10$ items, with indifference in the ranks of the remaining 8 items.

The paper is organized as follows. In section \ref{sec:cmm}, the Clustered Mallows Model is defined, and some conditions on $\boldsymbol{z}$ and $\theta$ are established so that the model is identifiable. We illustrate how the CMM is related to other approaches in the literature such as product partition models \citep{hartigan1990, BarryHartigan1992} and the Infinite Generalised Mallows (IGM) \citep{meila2010}. The CMM is defined in terms of distances between ordered clusters for which some options are given in section \ref{sec:distances}. The probability model involves a non-analytical normalisation and this poses challenges to statistical inference and model simulation. These points are handled in section \ref{sec:computation} where we present a pseudo-likelihood formulation and provide means to estimate the CMM. Fitting the CMM to observed ranks is handled in sections \ref{sec:bayes} and \ref{sec:greedy_search}. Bayesian inference for fitting the model parameters with some pre-specified structure, that is, $L$ and $n_1, \ldots, n_L$, is the focus of section \ref{sec:bayes}. In section \ref{sec:greedy_search}, an algorithmic search of the model structure is designed for when the goal is to learn both the partition and parameters from the data. Finally, real examples demonstrate the CMM fit in different scenarios. Formula 1 racing results are considered for analysis in section \ref{sec:formula1} with $L$ and $n_1, \ldots, n_L$ chosen to reflect characteristics of the competition's point system. We conclude with the analysis of the Tohoku sushi rankings in section \ref{sec:sushi} where the methods developed in sections \ref{sec:opt} and \ref{sec:bayes} are applied jointly. In both our real datasets, comparison to the Mallows model is explored, and modelling ordered clusters is demonstrated to be preferrable in each example. 


\section{The Clustered Mallows Model and related models}\label{sec:cmm}

Suppose that we have a collection of $q$ independent ranks from a population, 
\[\underline{\pi} = (\pi_1, \ldots, \pi_q), \; \pi_j \in \mathcal{P}_n \quad \forall\; j=1,\cdots,q.\] 
The Clustered Mallows model (CMM) places a probability distribution on $\underline{\pi}$ and depends on parameters $\boldsymbol{z} \equiv (z(1), \ldots, z(n))$ and $\theta>0$. Additionally it depends on a measure of agreement between ordered clusters, which is denoted by $d_{oc}(\cdot, \cdot)$ to make a distinction from (\ref{eq:mallows_model}). We provide more details on possible choices for $d_{oc}(\cdot, \cdot)$ in Section~\ref{sec:distances}. For now, let us write the CMM as


\begin{align}\label{eq:cmm}
f(\underline{\pi}| \boldsymbol{z}, \theta) = \prod_{j=1}^q \left( 
    \frac{\exp \left\{ -\theta d_{oc}(\pi_j ; \boldsymbol{z})  \right\}}{\Psi(\theta; \boldsymbol{z})} \right) =  \exp \left\{ -\theta \sum_{j=1}^q d_{oc}(\pi_j , \boldsymbol{z})  \right\} \Psi(\theta; \boldsymbol{z})^{-q},
\end{align}
where $\Psi(\theta; \boldsymbol{z}) = \sum_{\pi \in \mathcal{P}_n} \exp\{ -\theta d_{oc}(\pi, \boldsymbol{z}) \}$.

The main aim of modelling the observed collection $\underline{\pi}$ with (\ref{eq:cmm}) is to learn a consensus among the $q$ assessors and the variability around this, allowing for the potential that some items may be clustered and that the clusters themselves are ordered. 
In order to learn $\boldsymbol{z}$ and $\theta$ some conditions on $\boldsymbol{z}$ need to be established in order to ensure that the parameters are identifiable. This is because distances between ordered clusters behave differently in accordance with the number of clusters $L$, and items per cluster, $n_l(\boldsymbol{z}) = \sum_{i}^n I\{ z(i) = l \}$. As the number of exchangeable items varies in $\boldsymbol{z}$, so too does the distribution of $d_{oc}(\pi, \boldsymbol{z})$. For example, if $n_1(\boldsymbol{z}) = n$ and all items belong to the same cluster, the distance will be zero for any $\pi$. This is obviously not the case if $L>1$. It turns out that $L$ and $n(\boldsymbol{z})$ determine the distance distribution, something that will be made clear in section (\ref{sec:distances}) once $d_{oc}$ is specified. 

For now, let us term $n(\boldsymbol{z}) = (n_1(\boldsymbol{z}), \ldots, n_L(\boldsymbol{z}))$ the \textit{Clustering Table} (in short, CT). The identification of the CMM is established by fixing the CT, guaranteeing the interpretability of $\theta$, and achieving an important simplification of (\ref{eq:cmm}). Given the CT, $\Psi(\theta; \boldsymbol{z}) = \Psi(\theta)$ and the normalisation term does not depend on $\boldsymbol{z}$. This can be thought of as the CMM counterpart of the MM's right-invariance of $d(\cdot, \cdot)$, under which $\phi(\alpha, \pi_0) = \phi(\alpha)$. Nonetheless, $\Psi(\theta)$ is still non-trivial to compute, and this issue is addressed in section \ref{sec:normalisation}.

Fixing the CT means that the number of clusters and items per group are pre-specified. This requirement does not make the proposed model limited in practice, but the choice of the CT is treated as a model selection problem. Section \ref{sec:opt} is devoted to choosing the CT where an algorithmic search is outlined. 

The CMM probability model with the set CT $n(\boldsymbol{z}) = (n_1(\boldsymbol{z}), \ldots, n_L(\boldsymbol{z}))$ is  

\begin{equation}\label{eq:cmm_fixed}
    f(\underline{\pi}| \boldsymbol{z}, \theta) =  \frac{\exp \left\{ -\theta \sum_{j=1}^q  d_{oc}(\pi_j, \boldsymbol{z})  \right\}}{\Psi(\theta)^q}, \quad \theta >0, \boldsymbol{z} \in \frac{\mathcal{P}_n}{\mathcal{P}_{n_1}, \times \cdots \times \mathcal{P}_{n_L} },
\end{equation}
where $d_{oc}(\pi, \boldsymbol{z})$ is a comparison between the reference group $\{ \boldsymbol{z}\} = ( \{z^{-1}(1)\}, \ldots, \{z^{-1}(L)\} )$ and $(\{ \pi(1), \ldots, \pi(n_1) \},$ $ \{ \pi(n_1 +1), \ldots, \pi(n_1 + n_2) \}, \ldots, \{ \pi(n_1 + \ldots, n_{L-1} +1),\ldots, \pi(n) \})$. The latter splits $\pi$ into clusters of sizes $n_1, \ldots, n_L$, according to the CT, denoted by $\{ \pi \}_{CT}$ or simply $\{ \pi \}$. As an example, with $CT= (n_1, n_2, n_3)  = (2,2,1)$, $\boldsymbol{z} = (2,1,1,2,3)$ and $\pi =(2,1,3,4,5)$ the baseline group is $\{ \boldsymbol{z} \} = (\{2,3\}, \{1,4\}, \{5\})$ and the other is $\{\pi\}$ is $(\{2,1\}, \{ 3,4\}, \{5\})$. Possible ways to measure the disagreement between the two ordered clusters are presented in section (\ref{sec:distances}). Before going into that, some important connections of the CMM are discussed below.

The CMM shares a close connection to the tied Mallows model by \cite{critchlow}. Here tied ranks are defined as the process of partitioning $n$ elements into $L$ ordered clusters, similar to our approach here. The main focus of \cite{critchlow} is to extend popular distances for $\pi, \pi_0 \in \mathcal{P}_n$ to ties in either or both $\pi$ and $\pi_0$. These are adopted later by \cite{marden1995} for hypotheses testing. \cite{marden1995} develops an approach to test the randomness of a ranking, that is, $\mathcal{H}_0: \pi \sim \mbox{Uniform}(\mathcal{P}_n)$. The probability that $\pi$ is generated from the null hypothesis is based on its distance to a tied $\pi_0^* = (1, \ldots, 1)$ that assigns all items the same label. Availing of the extended metrics by \cite{critchlow},  likelihood ratio tests are conducted using an exponential family distribution with sufficient statistic $d(\pi, \pi_0^*)$.

To the best of our knowledge, no attention was devoted to the extended metrics by \cite{critchlow}. They are applied in the CMM as the possible forms of $d_{oc}$ where the focus is on ranking aggregation with non-strictness in the consensus of $\underline{\pi} = (\pi_1, \ldots, \pi_q)$.

Another important related class is \textit{product-partition models} introduced by \cite{hartigan1990}, \cite{BarryHartigan1992}. This is also an approach to partition $n$ elements in $L$ sets. Let $\mathcal{S}_l$ denotes the $l^{th}$ group for $l =1,\ldots, L$. Product partition models assume that each object belongs to a single set, i.e., $\mathcal{S}_l \cap \mathcal{S}_{l'} = \emptyset$ for any $l \neq l'$ and $ \cup_{l=1}^L \mathcal{S}_l = \{1,\ldots, n\}$, which is also in place for our model. However, the joint probability of $\mathcal{S}_1, \ldots, \mathcal{S}_L$ is written as $p(\mathcal{S}_1, \ldots, \mathcal{S}_L ) = L \prod_{l=1}^L c(\mathcal{S}_l)$, where $c(\mathcal{S}_l)>0$ is a measure of set cohesion. This means that product-partition models assume between-set independence, something that does not hold in a CMM once rankings carry relative information. For a review of Bayesian product partition models, we recommend \cite{bayesian_ppm}. 

The Infinite Generalised Mallows (IGM) model by \cite{meila2010} is another important connection, developed for top$-k$ rankings, $\widetilde{\pi}$. It assumes that the first $k$ items are ranked from an infinite set of items and $\widetilde{\pi}$ is modeled according to $\widetilde{d}_k(\widetilde{\pi}, \sigma)$. The latter is the Kendall distance to an infinite permutation $\sigma$. By $\widetilde{d}_k$ it is denoted that the comparison is done up until rank $k$, which is a form of truncation of $d_k$. The CMM is analogous to the IGM when the CT is $n_l =1$ for $l =1, \ldots, k$ and $n_{k+1} = (n-k)$. This is an alternative (finite) approach to the task that sets the first $k$ as single element clusters, and the final cluster contains the remaining $(n-k)$ items. 

The CMM overcomes a possible limitation of the IGM, which is mentioned by the authors in section 3.4. When the data does not contain enough information about the relative ranking of some item pair $i$ and $i'$ or $\bar{p}(i, i')=0.5$, then the optimal IGM consensus does not have a unique solution. In this case, it is necessary to acquire more samples, which might not always be feasible. The CMM is a simple solution to this issue that simply assigns $i, i'$ to the same cluster.

\section{Measures of cluster ordination}\label{sec:distances}

This section specifies and compares distances between ordered clusters which will be denoted as $d_{oc}$. Here we build on the work of \cite{critchlow} where metrics of association in $\mathcal{P}_n$ are formally generalised to the tied space. As previously introduced, comparisons in the CMM are between $\{ \boldsymbol{z}\}$ and $\{ \pi \}_{CT} \equiv \{ \pi \}$. These are ordered clusters of identical sizes, and CT is a \textit{tie pattern} (\cite{marden1995}). Hence, Critchlow's extended metrics are applicable to quantify the agreement between $\{ \boldsymbol{z}\}$ and $ \{ \pi \}$.

\subsection{Hamming}

The Hamming distance between permutations $\gamma, \omega \in \mathcal{P}_n$ involves rank-wise comparisons $I\{ \gamma(i) \neq \omega(i) \}$ for $i \in \{1, \ldots, n\}$. The indicator function $I\{ \gamma(i) \neq \omega(i) \}$ is zero if and only if $\gamma(i)$ and $\omega(i)$ take the same value. Otherwise, $I\{ \gamma(i) \neq \omega(i) \} = 1$. The Hamming distance is denoted by $d_h(\gamma, \omega)$, and is given by the sum $d_{h}(\gamma, \omega) = \sum_{i=1}^n I\{ \gamma(i) \neq \omega(i) \}$. This is the simplest association metric in $\mathcal{P}_n$, and it has been extended in Critchlow to $\gamma^*, \omega^*$ containing ties. The author demonstrates that the Hamming extension is straightforwardly $d_{oc}(\gamma, \omega) = \sum_{i=1}^n I\{\gamma^*(i) \neq \omega^*(i) \} =1$. In other words, imposing no disagreements between the repetitions yields a valid metric in the space of ties. In the CMM, it will be denoted as $d_{oc, h}(\pi, \boldsymbol{z})$, and is defined below.

\begin{definition} The Hamming ordered cluster distance between the ordered clusters $\{\boldsymbol{z}\}$ and $\{\pi\}$ is 

\begin{align}
     d_{oc,h}(\pi, \boldsymbol{z}) = \sum_{i=1}^n I\{ z(\pi(i)) \neq {\widetilde{z}}_i  \}.
\end{align}
The symbol $\widetilde{\cdot}$ indicates sorting in increasing order, so ${\widetilde{z}}_i$ is the $i^{th}$ element of $\widetilde{\boldsymbol{z}}$. By $z(\pi(i))$ it is meant that the group label in $\boldsymbol{z}$ is applied to $\pi(i)$.
\vspace{0.2cm}

\noindent \textcolor{blue}{Example:} Take the previous $\boldsymbol{z} = (2,1,1,2,3)$ and $\pi = (2,1,3,4,5)$. The sorted $\widetilde{\boldsymbol{z}}$ is $(1,1,2,2,3)$ and $z(\pi(1)), \ldots, z(\pi(5))$ is $(1,2,1,2,3)$. The Hamming distance is $d_{oc, h}(\pi, \boldsymbol{z}) = I\{1 \neq 1\} + I\{1 \neq 2\} + \cdots + I\{3 \neq 3\} $ = 2.
\end{definition}\label{def:dgo_h}


\subsection{Kendall}

The Kendall metric in $\mathcal{P}_n$, here $d_k(\gamma, \omega)$, compares the order relation of pairs $\{ \gamma(i), \gamma(j)\}$ and $\{\omega(i), \omega(j)\}$. For all $i\neq j,\; i,j =1,\ldots,n$ each of the following two cases $\gamma(i) > \gamma(j), \omega(i) < \omega(j)$ or $\gamma(i) < \gamma(j), \omega(i) > \omega(j)$ counts as a disagreement. A summation over all such disagreements of $\{i,j\}$ combinations gives the Kendall metric in $\mathcal{P}_n$. The extension of this idea to ties is developed in Critchlow, and we adopt it to introduce the Kendall ordered cluster distance. This metric is denoted by $d_{oc, k}$ and is defined as follows.

\begin{definition} The Kendall ordered cluster distance between $\{ \boldsymbol{z} \}$ and $\{ \pi \}$ is given by
\begin{align}
       d_{oc,k}(\pi, \boldsymbol{z}) = \sum_{i=1}^{L-1} \sum_{j=1}^L n_{ij} \left( \sum_{i'= i +1}^L \sum_{j' = 1}^j n_{i'j'} \right),
\end{align}
where $n_{ij}$ is the number of items placed in group $i$ in $\{ \pi \}$ and in group $j$ of $\{ \boldsymbol{z}\}$.
\vspace{0.2cm}

\noindent \textcolor{blue}{Example:} Consider $\boldsymbol{z} = (2,1,1,2,3)$ and $\pi = (2,1,3,4,5)$ we obtain $n_{11} = 1, n_{12} = 1, n_{13} = 0, n_{21} = 1,n_{22}= 1,n_{23}= 0, n_{31}= 0, n_{32}= 0, n_{33}= 1$. The Kendall ordered cluster distance is then $d_{oc, k}(\pi, \boldsymbol{z}) = n_{11}(n_{21} +n_{31}) + n_{12}(n_{21}+n_{22}+n_{31}+n_{32}) + n_{13}(n_{21}+n_{22}+n_{23}+n_{31}+n_{32}+n_{33}) + n_{21}n_{31} + n_{22}(n_{31}+ n_{32}) + n_{23}(n_{31}+ n_{32}+ n_{33}) = 1(1) + 1(2) + 0(3) + 1(0) + 1(0) + 0(1) = 3$.
\end{definition}\label{def:dgo_k}

\begin{figure}
    \centering
    \includegraphics[width = 0.65\linewidth]{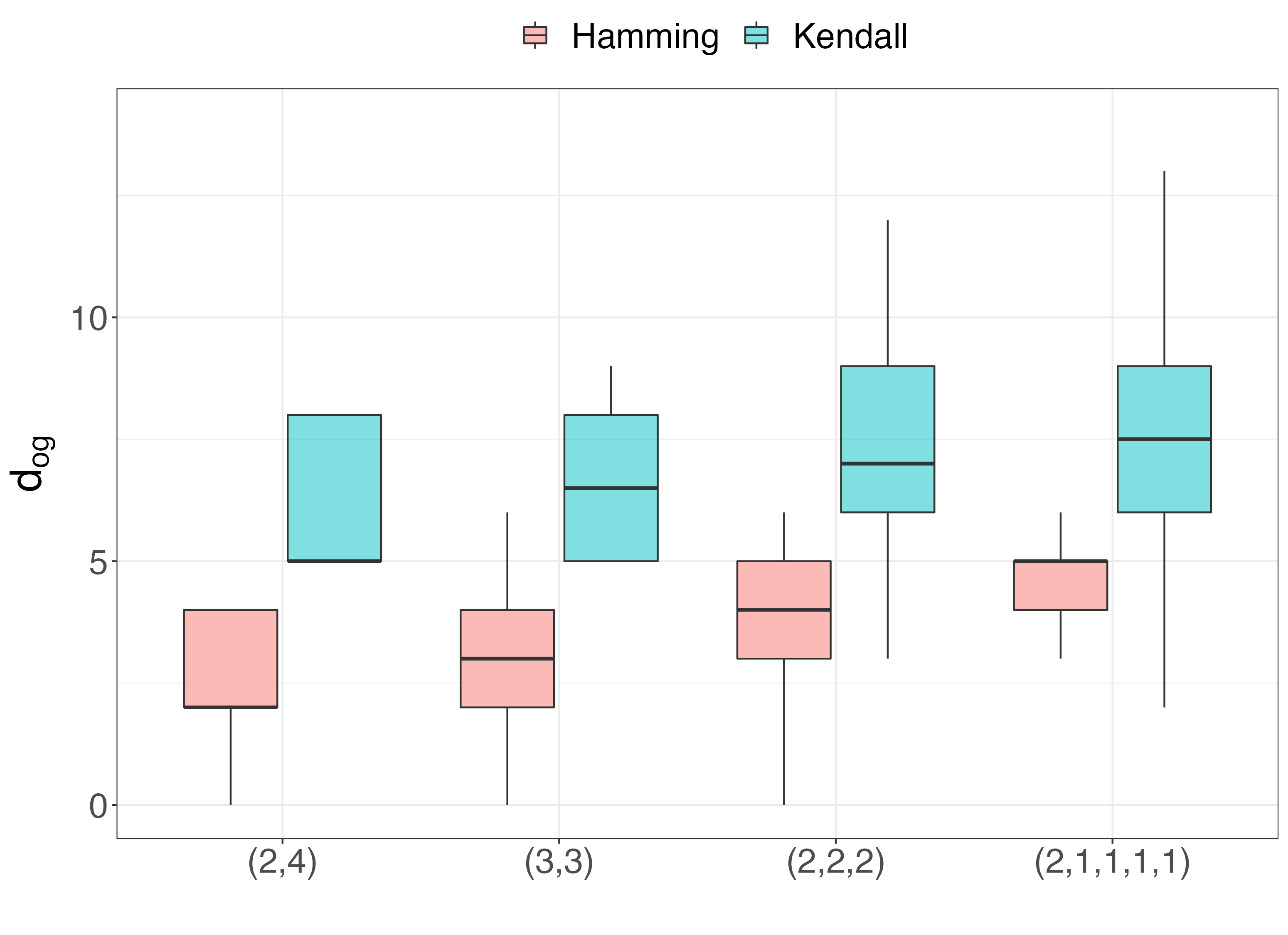}
    \caption{Distribution of the Hamming and Kendall ordered cluster distances $d_{oc}$'s under some example CTs that partition $n=6$ items.}
    \label{fig:distances_distribution}
\end{figure}

Figure (\ref{fig:distances_distribution}) illustrates the distribution of $d_{oc, h}$ and $d_{oc, k}$ obtained from different CT configurations.  As it was mentioned in section (\ref{sec:cmm}), the CT determines the disagreements that can be made, hence the $d_{oc}$ distribution. This is because the clustering table stipulates the indistinguishable items, and indifference does not contribute to $d_{oc}$. As to be expected, $d_{oc}$ values depend also on the type of comparison, so Figure (\ref{fig:distances_distribution}) shows the Hamming and Kendall cases for each CT. 

The next subsection is devoted to further illustrate and compare $d_{oc, h}(\pi, \boldsymbol{z})$ and $d_{oc, k}(\pi, \boldsymbol{z})$ when used in Equation (\ref{eq:cmm_fixed}) together with some parameter configurations to exemplify the Clustered Mallows Model.


\subsection{Distances properties}\label{sec:dists_prop}

The Kendall and Hamming ordered cluster distances are now illustrated under some configurations of $\boldsymbol{z}$ and $\theta$. Our aim is to provide a further understanding of their properties in the context of the CMM. We consider $\boldsymbol{z}_1 = (1,1,2,2,3,3)$ and $\boldsymbol{z}_2 = (1,1,1,2,2,3,3,3,4)$ which encode the ordered clusters $\{ \boldsymbol{z}_1 \} = ( \{1, 2\}, \{3, 4\}, \{5, 6\})$ and $\{ \boldsymbol{z}_2\} = (\{1,2,3\}, \{4,5\}, \{6,7,8\},\{9\} )$. Setting $\boldsymbol{z}_1$ has three equally sized clusters of $n=6$. In $\boldsymbol{z}_2$, $n=9$ items are 
partitioned into $L=4$ clusters. For this setting, there are three top choices (items $\{1,2,3\}$), $\{9\}$ is the worse alternative, and others are in two intermediate placings.

To study these settings alongside $d_{oc,h}, d_{oc, k}$, a useful summary is the aggregation by \textit{rank and cluster}. This is the probability that the model takes an element of cluster $l \in \{1, \ldots, L\}$ in each rank $i$. Definition \ref{rank_type_probs} formalises this concept, which we name the Rank Cluster (RC) probability.

\begin{definition}
The CMM$(\boldsymbol{z}, \theta)$ Rank-Cluster (RC) probability is 
\begin{equation} 
    RC(i, l) = P(z(\pi(i)) = l|\boldsymbol{z}, \theta) = \sum_{\pi \in \mathcal{P}_n} f(\pi|\boldsymbol{z}, \theta) I\{ z(\pi(i))=l\} = E_{f(\cdot|\boldsymbol{z}, \theta)}[I\{ z(\pi(i))=l\}]
     \end{equation}\label{rank_type_probs}
for $i \in \{1, \ldots, n\}$ and $l \in \{1, \ldots, L\}$. The indicator function $I\{ z(\pi(i))=l\}=1$, if $z(\pi(i))=l$ and zero, otherwise.
\end{definition}

\begin{figure}[H]
    \centering
    \includegraphics[width = 0.7\linewidth]{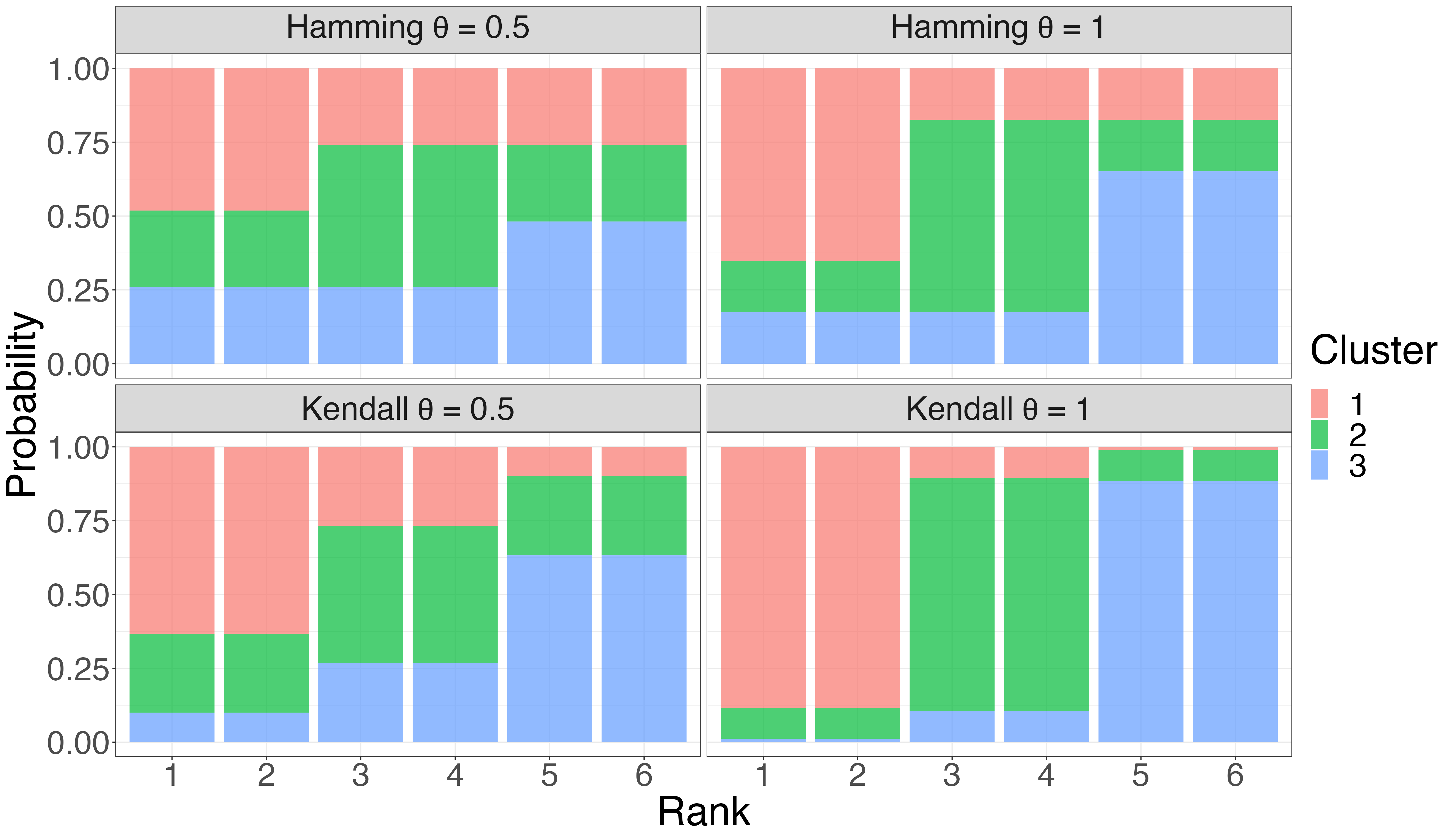}
    \caption{CMM rank-cluster probabilities under Hamming and Kendall ordered cluster distances with $\boldsymbol{z}= (1,1,2,2,3)$ and $\theta = 0.5$ on the left, $\theta =1$ on the right.}\label{fig_distances}
\end{figure}

We illustrate various RC probabilities using $\boldsymbol{z}_1$ as defined above and each of the two possible values $\theta = 0.5$ or $\theta = 1$ in Figure \ref{fig_distances}. In each window, bars correspond to ranks $i \in \{1, \ldots, 6\}$, colored according to the probabilities of $l \in \{1,2,3\}$. The first row is computed using Equation (\ref{rank_type_probs}) and $d_{GO, h}$, with $\theta = 0.5$ on the left and $\theta = 1$ on the right. The same values of $\theta$ are used when computing (\ref{rank_type_probs}) with $d_{GO, k}$ in the second row. The effect of increasing $\theta$ is to sharpen the probability of a \textit{correct cluster assignment} for all $i$ for either distance. Correct cluster assignments are those that preserve $\boldsymbol{\widetilde{z}_1} = (1,1,2,2,3,3)$, i.e., $z(\pi(1)), \ldots, z(\pi(6)) = \boldsymbol{\widetilde{z}_1}$. The spread around $\widetilde{\boldsymbol{z}}_1$ is reduced as $\theta \rightarrow \infty$, shown by increased red bars for ranks 1 and 2, green for $i \in \{3,4\}$, and blue for $i \in \{5,6\}$.
 
In addition to the effect of $\theta$, Figure \ref{fig_distances} illustrates that the clustering in  $\boldsymbol{z}$ translates to RC. If $i, i'$ are ranks that expect the same correct cluster, i.e. $\widetilde{z}_i = \widetilde{z}_{i'}$, then $RC(i, l) = RC(i', l) \quad \forall \; l=1,\dots,L$. For example, the first and second bars ($i=1, i' = 2$) are identical in all windows since $\widetilde{z}_1 = \widetilde{z}_2 = 1$. On the contrary, $i=1, i' = 3$ are different once $\widetilde{z}_1 = 1, \widetilde{z}_2 = 2$. We expand on this property in the next remark.

\begin{remark} 
If $j \neq j'$ are tied items from the set $\{1, \ldots, n\}$, then $z(j) = z(j') = l$ for some $l$ in $\{1, \ldots, L\}$. Indifference between $j$ and $j'$ immediately implies that they have the same probability of being ranked $i^{th}$. Mathematically $P( \pi^{-1}(i) = j) = P( \pi^{-1}(i) = j')$ if $z(j) = z(j')$. Rank exchangeability is an implication of item-indifference which can be expressed as $P(z(\pi(i)) = l) = P(z(\pi(i')) = l) \; \forall \; l \mbox{ if } \widetilde{z}_i = \widetilde{z}_i'.$
\end{remark}\label{remark_diff_distances}

\begin{figure}[H]
    \centering
    \includegraphics[width = 0.7\linewidth]{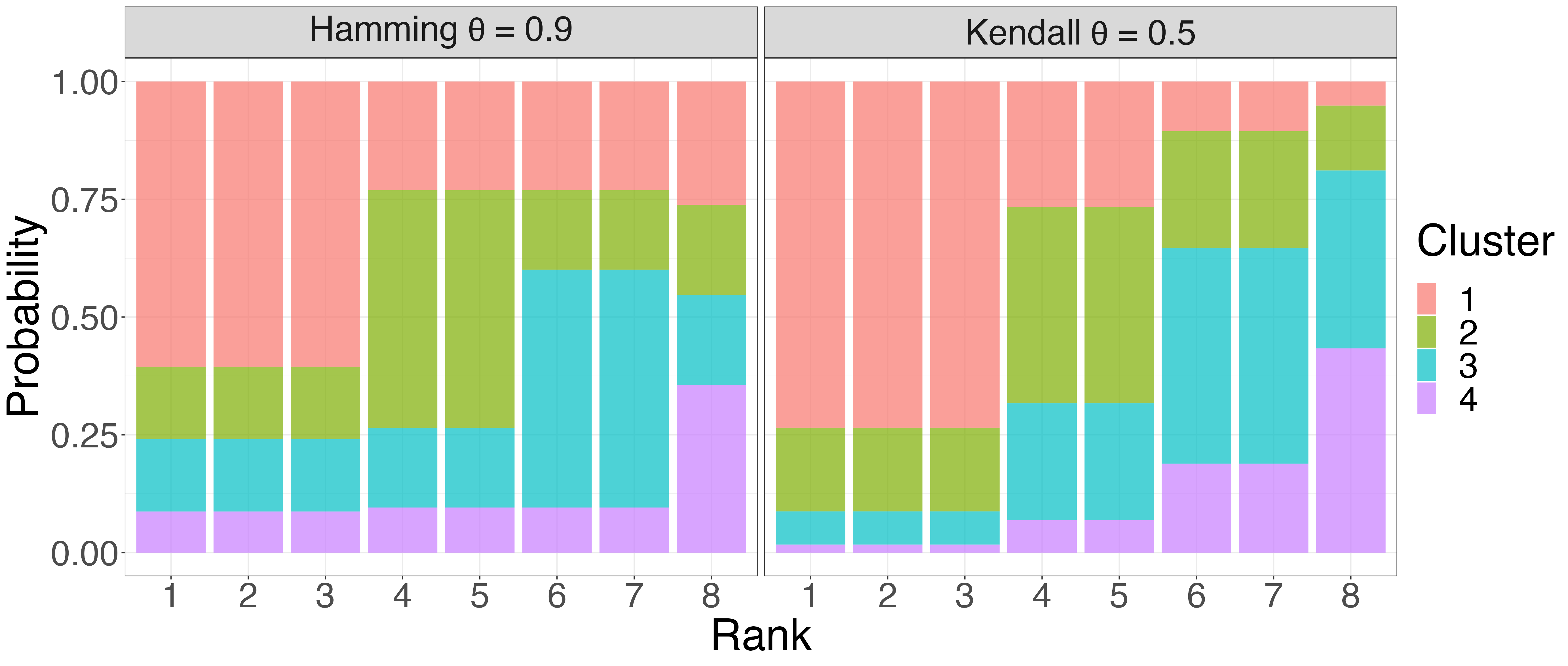}

        \caption{Rank-cluster probabilities of the Hamming ordered cluster distance with $\theta = 0.9$ on the left, and Kendall ordered cluster distance with $\theta = 0.5$ on the right. The cluster allocations are $\boldsymbol{z}_2 = (1,1,1,2,2,3,3,3,4)$ in this example.     }\label{fig_distances2}
\end{figure}

Similar behaviour is observed in Figure \ref{fig_distances2}, which focuses on $\boldsymbol{z}_2$. This configuration has the clustering table $(3,2,3,1)$ which shows the equivalence of the $RC$'s $\{1,2,3\}$,  $\{4,5\}$ and $\{6,7\}$. The Hamming distance and $\theta =0.9$ are used to obtain the RC values to the left of Figure \ref{fig_distances2}. On the right, $d_{oc}$ corresponds to the Kendall distance and $\theta$ is $0.5$. Figure \ref{fig_distances2} shows that the Kendall ordered cluster distance is more around $\widetilde{\boldsymbol{z}}_2$ in comparison to the Hamming ordered cluster distance, even with the smaller $\theta =0.5$. 

It is immediate from the definitions that $d_{oc, k}(\pi, \boldsymbol{z}) \geq d_{oc, h}(\pi, \boldsymbol{z})$ for fixed $\pi$. Making $\theta$ values comparable could be achieved with normalisation. When there are no ties, $d_{h}$ takes values in $[0, n]$ and $d_k \in [0, {n \choose 2}]$. However, there is no expression for $\max_{\pi \in \mathcal{P}_n} d_{oc,h}(\pi,\boldsymbol{z})$ or $\max_{\pi \in \mathcal{P}_n} d_{oc,k}(\pi,\boldsymbol{z})$. See \cite{critchlow} for further details. The consequence is that $\theta$ will not be comparable between CMMs taking different distances. This is a remark that should be borne in mind for the remainder of the paper.


\section{Computational aspects}\label{sec:computation}

The ordered clusters distances presented in section \ref{sec:distances} complete the specification of the CMM. We are now ready to tackle aspects such as model simulation and inference. In section \ref{sec:simulation}, we handle the sampling task, where the goal is to produce random realisations from a CMM with specified $\theta, \boldsymbol{z}$, $d_{oc}$ and CT. Subsequently, section \ref{sec:normalisation} details the model's normalisation constant $\Psi(\theta)$, which is non-trivial to compute.

\subsection{Model simulation}\label{sec:simulation}

Sampling from distance-based probability models is often a non-trivial task. In the Mallows model, approaches to draw exactly from (\ref{eq:mallows_model}) are limited to some distance choices and moderate $n$ values. One example is the \textit{distances algorithm} \citep{Irurozki2019}, which requires knowing the number of permutations at each distance value. Most often, $\pi$ is sampled approximately, using MCMC methods. 


MCMC sampling $\pi \in \mathcal{P}_n$ often follows the strategy proposed by \cite{diaconis2009}. In this algorithm, a candidate state $\pi'$ is generated from $\pi$ via a random transposition. The transposition is between a pair $\{i, j\}$ where $i, j \in \{1, \ldots, n\}, i\neq j$ are drawn uniformly at random. The proposed state $\pi'$ results from $\pi$ by swapping $\pi(i)$ with $\pi(j)$ and vice versa. This is then accepted or rejected with the usual Metropolis probability. 

It has been shown that this simple and computationally efficient method converges quickly to the target stationary distribution. The guideline proposed by \cite{diaconis2009} is that $n \log n$ steps suffice to reach stationarity, approximately. 
This strategy is widely applied to sample the MM, and is often an efficient sampler. It extends directly to a sample from a CMM with parameters $\boldsymbol{z}$, $\theta$ and some $d_{oc}$. This is described in Algorithm \ref{alg:cmm_sampler}. A Markov chain with stationarity distribution $f(\cdot|\boldsymbol{z}, \theta)$ is constructed using the pairwise transpositions proposal. The last state after $N$ iterations of the algorithm is taken as an approximate draw from the CMM. Naturally, the initial $\boldsymbol{z}$ conveys the model's CT.

\begin{algorithm}
\DontPrintSemicolon
\SetAlgoLined
\SetNoFillComment

\KwIn{$N:$ number of iterations; initial $\pi$; model parameters $\theta$, $\boldsymbol{z}$, $d_{oc}$ choice }
\For{$i \in 1:N$}{

Sample $j,k$ from $\{1,\ldots,n\}$ without replacement;

Set $\pi' \leftarrow \pi$

Switch positions $j$ and $k$: $\pi'(j) \leftarrow \pi(k)$ and $\pi'(k) \leftarrow \pi(j)$;

Compute $\alpha(\pi')  = \min \left\{ 1, \frac{f(\pi'|\boldsymbol{z}, \theta)}{f(\pi|\boldsymbol{z}, \theta)}  =  \frac{\exp(-d_{oc}(\pi',\boldsymbol{z}))\bcancel{\Psi(\theta)}}{\exp(-d_{oc}(\pi,\boldsymbol{z}))\bcancel{\Psi(\theta)}} \right\}$;

With probability $\alpha$, set $\pi \leftarrow \pi'$;

}
\textbf{Output:} $\pi$
\caption{Drawing from a CMM probability model using $N$ iterations of a Metropolis algorithm. A symmetric pairwise switch proposal generates candidate states, and the last iteration of the chain is approximately a CMM($\boldsymbol{z}, \theta$) draw.}
\end{algorithm}\label{alg:cmm_sampler}

Collecting $q$ independent CMM samples can be achieved with a parallel implementation of Algorithm (\ref{alg:cmm_sampler}), and this will be denoted by the function \textcolor{blue}{$rCMM(\boldsymbol{z}, \theta, N, q)$}. Note that the intractable normalisation term, $\Psi(\theta)$, cancels in the Metropolis acceptance probability $\alpha(\pi')$. 


\subsection{Pseudolikelihood approximation}\label{sec:normalisation}




This section is devoted to the problem of computing the CMM probability, $f(\cdot|\boldsymbol{z}, \theta)$ for observed $\pi$. The latter depends on the constant $\Psi(\theta) = \sum_{\pi \in \mathcal{P}_n} \exp\{ -\theta d_{oc}(\pi, \boldsymbol{z}) $ which is generally intractable. To address this, we formulate a \textit{pseudo-likelihood} model, which can be used in different ways. One immediate usage is to simply replace the true likelihood and carry out approximate inference with this tractable counterpart. Another direction is to gather pseudo-likelihood samples to construct a statistical estimator of $f(\pi|\boldsymbol{z}, \theta)$. 




Let $\widetilde{f}(\cdot|\boldsymbol{z}, \theta)$ denote a 
\textit{multistage} approximation of the true $f(\cdot|\boldsymbol{z}, \theta)$, our pseudo-likelihood model. By multistage, we mean that $\pi$ is formed with a sequence of $n-1$ independent decisions as explained next. Starting from the complete set of items, a ranking starts with the choice of $\pi(1)$. Once this is determined, $\pi(2)$ is a decision on top of $\{1, \ldots, n\}_{-\pi^{-1}(1)}$ where $-\pi^{-1}(1)$ denotes that $\pi^{-1}(1)$ has been removed from the possible alternatives. This goes on sequentially for $\pi(2), \ldots, \pi(n-1)$, until a single item is left for $\pi(n)$. The process of defining $\pi$ through sequential selections originating from top-ranking positions is known as \textit{forward ranking}.

Let $i = 1, \ldots, n-1$ index the $n-1$ choices and $\sigma_i$ denote the items available (unranked) at step $i$. Starting with $\sigma_1 = \{1, \ldots, n\}$, we approximate the conditional distribution $p(\pi(i)|\boldsymbol{z}, \theta, \sigma_i)$ for $i =1, \ldots, n-1$ as follows. First, a \textit{cluster decision} chooses the group label $z(\pi(i))$ of the $i-{th}$ ranked object. Then, the \textit{item stage} selects $\pi(i)$ conditionally on $z(\pi(i))$. In other words, the cluster step defines the ordered group that occupies position $i$ and given $z(\pi(i))=l$, the item $\pi(i)$ is drawn uniformly at random from the set of unranked label $l$ elements. 

The cluster stage works with an approximation of the rank-cluster probabilities $RC(i, l)$ (\ref{rank_type_probs}) that will be described in what follows. Let $CT_i$ denote the count of available items at step $i$ for labels $i=1, \ldots, L$. From $CT_i$, we can approximate the $RC(i,l)$ probability as follows. We introduce the term $D(l|CT_i)$, which is an indicator function of $CT_i$. The purpose of $D(l|CT_i)$ is to approximate the number of disagreements that can be made if a cluster $l$ item is in $\pi(i)$. This is done by setting $D(l|CT_i) = 0$ if $l$ is the smallest label in the unranked set, i.e. $n_l(\sigma_i) >0$. Otherwise, $D(l|CT_i) = n_l(z(\sigma_i))$. The approximate probability $\widetilde{RC}(i,l)$ is obtained from normalisation of $\exp\{ \theta D(l|CT_i) \}$, given by

\begin{equation}\label{pseudo_p}
\widetilde{RC}(i,l) = \frac{\exp\{-\theta D(l |CT_i)      \}}{\sum_{l=1}^{L} \exp\{-\theta D(l |CT_i ) \}}.
\end{equation}

To draw from the CMM pseudo-likelihood, the first step is then to sample $z(\pi(i))$ from $\{1, \ldots, L\}$ with probabilities $\widetilde{RC}(i,1),$ $\ldots, \widetilde{RC}(i,L)$. Or, equivalently, gather (\ref{pseudo_p}) for some observed $\pi$ from $z(\pi(i))$. Given $z(\pi(i)) =l$, the item probability is simply $1/\sum CT_i$, in line with the model assumption of within-group equivalence. To sample from $\widetilde{f}(\cdot|\boldsymbol{z}, \theta)$, an item label is then uniformly drawn from cluster $l$ objects that are currently unranked. After that, it is removed from $\sigma_{i+1}$ which is the available set at the next stage.

With this strategy, the probability $\widetilde{f}(\pi(i)|\sigma_i, \boldsymbol{z}, \theta)$ becomes a product of the two (\textit{cluster} and \textit{item}) steps, $\widetilde{f}(\pi(i)|\sigma_i, \boldsymbol{z}, \theta)  = \widetilde{RC}(i,l)(1/n_l(z(\sigma_i)))$, where $n_l = \sum_{j=1}^{\# \sigma_i} I\{ z(\sigma_i^j) = l\}$. Conditional independence of choices $i =1, \ldots, n-1$ gives us that the CMM pseudo-likelihood model $\widetilde{f}(\pi|\boldsymbol{z}, \theta)$ is

\begin{equation}
\widetilde{f}(\pi|\boldsymbol{z}, \theta) = \prod_{i=1}^{n-1} \widetilde{RC}(i,l)\frac{1}{ n_l(z(\sigma_i)) }.
\end{equation}\label{eq:f_tilde}

Computation of $\widetilde{f}(\pi|\boldsymbol{z}, \theta)$ in a small example where $n=4$ is illustrated next for clarity. Three conditionally independent choices are made on decreasing item sets starting from the first rank. Each choice is formed by the two-stage procedure that first selects a cluster label, and then an item in that group.

\textit{\noindent \textcolor{blue}{Example:} Let us consider $\boldsymbol{z} = (1,1,2,2)$ and $\theta = 0.5$. We work out the probability $\widehat{f}(\pi = (1,3,2,4)|\boldsymbol{z}, \theta)$ in Table \ref{tab:example_approximation}. The step-specific set of available items $\sigma_i$ and its clustering table are listed in the first two columns. The $CT_i$ is transformed into $D(l=1,2|CT_i)$ in column 3. This term assumes 0 for the smallest $l$, and $n_l(z(\sigma_i))$ otherwise, favouring the ordination of cluster labels. The $D(l=1,2|CT_i)$ is used to compute (\ref{pseudo_p}), and the product $\widetilde{RC}(i,l)(1/n_l(z(\sigma_i)))$ is $\widetilde{f}(\pi(i)|\sigma_i, \boldsymbol{z}, \theta)$. The final value of $\widetilde{f}(\pi|\boldsymbol{z}, \theta)$ is the product in the fifth column, in the logarithm scale, $\log \widetilde{f}(\pi = (1,3,2,4)|\boldsymbol{z}= (1,1,2,2), \theta =0.5) = -3.487$. The CMM probability for this setting is $-3.344$ with the Hamming ordered cluster distance, and $-3.593$ with the Kendall ordered cluster distance.}

\begin{table}[]
\scriptsize
\centering
\begin{tabular}{@{}cccccc@{}}
\toprule
$i$ & $\sigma_i$ & $CT_i$ & $D(l=1,2|CT_i)$ & $\widetilde{RC}(i, 1=1,2)$ & $\widetilde{f}(\pi(i)|\sigma_i, \boldsymbol{z}, \theta)$ \\ \midrule
1   & (1,2,3,4)  & (2,2)  & (0,2)           & (0.731, 0.269)             & 0.731 $\times$ 0.5                                                                                                                              \\
2   & (3,2,4)    & (1,2)  & (0,2)           & (0.731, 0.269)             & 0.269 $\times$ 0.5                                                                                                                              \\
3   & (2,4)      & (1,1)  & (0,1)           & (0.622, 0.378)              & 0.622 $\times$ 1                                                                                                                                \\ \bottomrule
\end{tabular}
\caption{Probability of $\pi = (1,3,2,4)$ computed from the CMM multistage approximation, $\widetilde{f}(\pi|\boldsymbol{z}, \theta)$.}\label{tab:example_approximation}
\end{table}

An importance sampling estimator of $\Psi(\theta)$ can now be computed using independent draws from $\widetilde{f}(\pi|\boldsymbol{z}, \theta)$. 
In this strategy, the CMM pseudo-likelihood acts as the importance density, it being a distribution that resembles the target, $f(\cdot|\boldsymbol{z}, \theta)$. The importance sampling estimator of $\widehat{\Psi}(\theta)$ based on $M$ pseudo-likelihood draws is given by

\begin{equation}\label{eq:Psi_IS}
    \widehat{\Psi}(\theta) = \frac{1}{M} \sum_{j=1}^M \frac{ \exp \left( -\theta d_{oc}(\pi_j,\boldsymbol{z}) \right)}{  \widetilde{f}(\pi_j|\boldsymbol{z}, \theta)}.
\end{equation}

Evidently, the true value of $\Psi(\theta)$ is a function of $n$ and $\theta$ which grows as $n \rightarrow \infty$ and $\theta \rightarrow 0$. In addition, it is important to note that the variability of $\widehat{\Psi}(\theta)$ is sensitive to the choice of $M$. With the importance density (\ref{eq:f_tilde}), we have that $\mbox{Var}(\Psi(\theta)) = E_{\widetilde{f}}[(w(\pi) - 1)^2]/M$, where $w(\pi) = \widetilde{f}(\pi|\boldsymbol{z}, \theta)/f(\pi| \boldsymbol{z}, \theta)$. This implies that high $M$ may be required depending on the precision to which $\widehat{\Psi}(\theta)$ must be computed. In the applications of (\ref{eq:Psi_IS}) explored in this paper, replication of $\widehat{\Psi}(\theta)$ with fixed $M$ is used to account for its variability. Our approach shares similarities to the methodology in \cite{vitelli2018}, where importance sampling with a pseudo-likelihood has been used successfully for the Mallows distribution. Their likelihood approximation is also based on a forward-ranking multistage formulation, used in IS to estimate the Mallow's normalisation for distances where this term does not have a closed form. Naturally, their methodology is not directly applicable so we have designed a pseudo-likelihood that embeds the CMM clustering structure.

If the computational cost related to this procedure becomes prohibitive, one option is to use $\widetilde{f}(\pi|\boldsymbol{z}, \theta)$ as an approximation of the true ${f}(\pi|\boldsymbol{z}, \theta)$. Inference with approximate (or composite) likelihoods has been widely applied to problems involving complex joint probabilities. When the underlying multivariate probability model is cumbersome to evaluate, composite likelihood methods adopt simplifications of the underlying dependency structure. Composite likelihood ideas can be used for fast approximate inference with the CMM, and we recommend \cite{varin2011overview} for more details on the topic.


\subsubsection{Model selection}\label{sec:info_criteria}

CMM model selection is the task of choosing the CT that best fits the observed data, for a specified distance $d_{oc}$. A common strategy in this regard is to use criteria that take the optimized log-likelihood and applies some penalisation for complexity. For instance, the Bayesian Information Criterion (BIC) for a CMM model $\mathcal{M}$ is $BIC(\mathcal{M}) = k\log(q) -2 f(\underline{\pi}|\boldsymbol{\widehat{z}}, \widehat{\theta})$, where $\{ \boldsymbol{\widehat{z}}, \widehat{\theta} \}$ are the parameter values that maximize $f$, and $k$ is the number of model parameters. The BIC penalisation, however, is not immediate for a CMM because the number of free parameters $k$ is not defined in the usual sense. In our context, the size of $\boldsymbol{z}$ is always $n$, and what changes is the number of repeated elements. 

We design a suitable penalisation that is based on the possible distinct arrangements of $\boldsymbol{z}$ for a given CT. With $n$ items, the least parameterized model possible is $CT =n$, the Uniform distribution on $\mathcal{P}_n$. At the other end is the Mallows Model, where $n_l =1 \; \forall \; l=1,\dots,L$, where $L=n$, and each item is its own cluster. The number of possible arrangements are, respectively, $1$ and $n!$ and, if configurations are equally likely, the probability $p(\boldsymbol{z}|CT)$ is $(n!/(n_1!\times n_L!))^{-1}$. This can be seen as an uninformative prior distribution for $\boldsymbol{z}$ constricted on the CT. If this is a prior distribution of $\boldsymbol{z}$, then the joint probability $f(\underline{\pi}, \boldsymbol{z}|\theta) = f(\underline{\pi}|\boldsymbol{z}, \theta)p(\boldsymbol{z}|n(\boldsymbol{z}))$ is automatically penalised for complexity. The CMM selection criterion that arises from this observation is then

\begin{equation}\label{eq:info_criteria}
    \widehat{\mathcal{I}}(CT) = \log  \widehat{f}(\underline{\pi}|\boldsymbol{\widehat{z}}, \widehat{\theta}) + \log p(\widehat{\boldsymbol{z}}|n(\boldsymbol{z})),
\end{equation}
where the likelihood is estimated using (\ref{eq:Psi_IS}). The above formulation has similarities with the Integrated Complete Likelihood (ICL, \cite{ICL}) criterion which is well cemented in model-based clustering. The latter is based on the idea of maximizing the joint probability of a mixture and the allocations vector, which is in some resemblance with $f(\underline{\pi}, \boldsymbol{z}|\theta)$.


With the ingredients just outlined, we are ready to develop inference with the CMM and an observed collection of ranks. This is done by assuming that $\underline{\pi}$ is a realisation of model (\ref{eq:cmm}) and the goal is to learn about $\boldsymbol{z}$ and $\theta$. Section \ref{sec:bayes}
handles Bayesian inference for $\boldsymbol{z}$ and $\theta$ given a model structure, CT. Algorithms to sample the doubly-intractable posterior are specified and we also handle incomplete observations of $\underline{\pi}$ via data augmentation. Strategies for model structure evaluation follow in section \ref{sec:greedy_search}. An algorithm is designed to tackle the CT search, which is useful when the goal is to learn both the model structure and parameters from observed data. 



\section{Inference: model parameters}\label{sec:bayes}

In this section, our focus is on the posterior distribution of $\boldsymbol{z}, \theta$ given the data, $\underline{\boldsymbol{\pi}}$,  
\begin{equation}\label{eq:cmm_posterior}
    p(\boldsymbol{z}, \theta|{\underline{\pi}}) \propto f(\underline{\pi}|\boldsymbol{z}, \theta) p(\theta) p(\boldsymbol{z}) = \exp\left\{ - \theta \sum_{j=1}^q d_{oc}(\pi_j, \boldsymbol{z}) \right\} \Psi(\theta)^{-q} p(\theta) p(\boldsymbol{z}),
\end{equation}
where $p(\boldsymbol{z})$ and $p(\theta)$ are prior distributions for $\boldsymbol{z}$ and $\theta$, respectively. Our strategy is to use MCMC sampling to infer (\ref{eq:cmm_posterior}). To this end, a Markov chain that has stationary distribution (\ref{eq:cmm_posterior}) is constructed. 
We do this by drawing from the full-conditional distributions of $\theta$ and $\boldsymbol{z}$. Algorithms to sample $p(\boldsymbol{z}|\underline{\pi}, \theta)$ and $p(\theta|\underline{\pi}, \boldsymbol{z})$ are outlined next.

The full-conditional distribution for $\boldsymbol{z}$ is written as 
\begin{equation}\label{eq:z_full_conditional}
    p(\boldsymbol{z}|\underline{\pi}, \theta) \propto \exp\left\{ -\theta \sum_{j=1}^q d_{oc}(\pi_j, \boldsymbol{z})\right\} p(\boldsymbol{z}),
\end{equation}
and does not involve the intractable constant $\Psi(\theta)$. Hence, standard MCMC techniques are suitable to sample (\ref{eq:z_full_conditional}). Metropolis moves that propose candidate states from random switches are one possible strategy. From initial states $\boldsymbol{z}^t$ and $\theta^t$, the proposal denoted as $\texttt{switch}(\boldsymbol{z}^t, m)$ randomly switches $m$ elements of $\boldsymbol{z}$. This is equivalent to the strategy employed in Algorithm (\ref{alg:cmm_sampler}), where $m=2$. The algorithm accepts $\boldsymbol{z}'$ as its next state with the Metropolis probability $\alpha(\boldsymbol{z}')$ which is  $$\alpha(\boldsymbol{z}') = \exp\left\{ -\theta^t \left(\sum_{j=1}^q d_{oc}(\pi_j,\boldsymbol{z}^{'}) - \sum_{j=1}^q d_{oc}(\pi_j,\boldsymbol{z}^t) \right) \right\} \frac{p(\boldsymbol{z}')}{p(\boldsymbol{z}^t)}.$$

Given $\boldsymbol{z}^{t+1}$, the next step is to sample $\theta^{t+1}$ from $p(\theta|\underline{\pi}, \boldsymbol{z}^{t+1})$. This is given by $p(\theta|\underline{\pi}, \boldsymbol{z}^{t+1}) \propto f(\underline{\pi}|\boldsymbol{z}, \theta) p(\theta) = \exp\left\{ -\theta \sum_{j=1}^q d_{oc}(\pi_j, \boldsymbol{z}^{t+1})\right\} \Psi(\theta)^{-q} p(\theta)$. Unavailability of $\Psi(\theta)$ makes sampling $p(\theta|\underline{\pi}, \boldsymbol{z}^{t+1})$ a non-trivial task. For instance, a Metropolis-Hastings (MH) algorithm requires evaluating $\Psi(\theta^t)/\Psi(\theta')$ in the acceptance probability. Specialised methodology has been developed for MCMC sampling in this context where the likelihood involves some non-analytical term. The problems in this class are often called \textit{doubly-intractable} (\cite{murray2006}) from likelihood intractability in addition to the model evidence. Some possible approaches are pseudo-marginal MCMC \citep{andrieu2009}, the Exchange Algorithm (EA) \citep{murray2006} and the noisy Metropolis algorithm \citep{alquier16}.

We adopt a version of the EA to sample $p(\theta|\underline{\pi}, \boldsymbol{z}^{t+1})$ using tools presented in Section \ref{sec:computation}. In EA, cancellation of intractable terms is achieved in the MH ratio using an augmentation strategy.
The target posterior is augmented with auxiliary draws from $f$ and the joint acceptance of $\theta' \sim h(\cdot|\theta^t)$ and $\underline{\pi}' \sim f(\cdot|\theta', \boldsymbol{z}^{t+1})$ is evaluated. Writing $f(\cdot|\theta, \boldsymbol{z})$ as $q(\underline{\pi}|\theta, \boldsymbol{z})/\Psi(\theta)$ where $q(\underline{\pi}|\theta, \boldsymbol{z}) = \exp\{ -\theta  \sum_{j=1}^q d_{oc}(\pi_j, \boldsymbol{z})\}$ allows us to see that the EA acceptance ratio is tractable. It is given by

$$\alpha(\theta') = \min \left\{\dfrac{p(\theta') q(\underline{\pi}|\theta', \boldsymbol{z}^t) q(\underline{\pi}'|\theta^t, \boldsymbol{z}^t) h(\theta^t|\theta') \bcancel{\Psi(\theta^t)^q} \bcancel{\Psi(\theta')^q}}{p(\theta^t) q(\underline{\pi}|\theta^t, \boldsymbol{z}^t) q(\underline{\pi}'|\theta', \boldsymbol{z}^t)  h(\theta'|\theta^t) \bcancel{\Psi(\theta')^q} \bcancel{\Psi(\theta^t)^q}}\right\}.$$\label{eq:ea_acceptance} 

With this clever augmentation choice, the unavailable terms cancel in $\alpha(\theta')$. The EA steps just described are summarised in Algorithm \ref{alg:exchange}. Therein, STEP 2 calls the sampling routine $rCMM(\boldsymbol{z}, \theta, N, q)$ to obtain the auxiliary data. Since $\underline{\pi}'$ is an approximate draw from a Gibbs run, Algorithm (\ref{eq:ea_acceptance}) is a \textit{Approximate Exchange} update.  The Approximate Exchange Algorithm (AEA) was introduced in \cite{caimofriel2011} and is an important extension of the EA that originally requires exact samples.

\begin{algorithm}

\KwIn{\mbox{Current } $\theta^{t}, \boldsymbol{z}^{t}$ and observed data ${\underline{\pi}}$}
 
\textcolor{gray}{STEP 1:} Draw $\theta'$ from some proposal distribution,  $\theta' \sim h(\cdot|\theta)$;

\textcolor{gray}{STEP 2:} Sample auxiliary data $\underline{\pi}'$ from the CMM using the \textcolor{blue}{$rCMM(\boldsymbol{z}, \theta, N, q)$} routine; 

\textcolor{gray}{STEP 3:} Accept $\theta^{t+1} \leftarrow \theta'$ with probability
$$\alpha(\theta') = \min \left\{1, \dfrac{p(\theta') q(\underline{\pi}|\theta', \boldsymbol{z}^t) q(\underline{\pi}'|\theta^t, \boldsymbol{z}^t) h(\theta^t|\theta')  }{p(\theta^t) q(\underline{\pi}|\theta^t, \boldsymbol{z}^t) q(\underline{\pi}'|\theta', \boldsymbol{z}^t)  h(\theta'|\theta^t) }\right\}. \nonumber$$
otherwise $\theta^{t+1} \leftarrow \theta^t$;

\caption{ \textcolor{blue}{$AEA(\theta^t, \boldsymbol{z}^t, N)$} Approximate Exchange Algorithm for sampling the full-conditional distribution of $\theta$.}\label{alg:exchange}
\end{algorithm}

With the Metropolis and AEA moves, Bayesian inference is the result of running the Markov chain that updates in turn $\boldsymbol{z}^{t+1} \sim p(\cdot|\underline{\pi}, \theta)$ and $\theta^{t+1} \sim p(\cdot|\underline{\pi}, \boldsymbol{z}^{t+1})$. Running the chain for a long enough number of iterations, we can collect draws and treat then as realisations from the posterior (\ref{eq:cmm_posterior}).


\subsection{Partial observations}\label{sec:partial_ranks}

Partial observation of ranks is easily accommodated in the Bayesian approach with data augmentation. An incomplete observation of $\pi$, denoted by $\widetilde{\pi}$, can come from settings such as top$-k$ elicitation or pairwise preferences. In any of such cases, we will assume that $\widetilde{\pi}$ reflects incomplete information about $n$ alternatives and embeds the observed piece. For example in top$-k$ data, one observes $\widetilde{\pi}(1), \ldots, \widetilde{\pi}(k)$, however, $\widetilde{\pi}(k+1), \ldots, \widetilde{\pi}(n)$ are not ranked, and we consider them as missing.

With a collection of partially observed ranks $\widetilde{\underline{\pi}} \equiv (\widetilde{\pi}_1, \ldots, \widetilde{\pi}_q)$, we can treat $\pi_j$'s as unobserved random variables with some distribution $p(\pi_j)$. A uniform at random arrangement of the unranked items is a reasonable assumption in the absence of additional information. Data augmentation brings $\widetilde{\underline{\pi}}$ to ${\underline{\pi}}$ and the target posterior becomes $p(\underline{\pi}, \boldsymbol{z}, \theta|\widetilde{\underline{\pi}})$. 

From an initial set $\underline{\pi}^{t}$, the complete data update proposes each $\pi_j'$ by randomly permuting the unranked items to fill the missing positions of $\widetilde{\pi}_j$. The probability to accept $\pi_j^{t+1}$ as $\pi_j'$ is $\alpha(\pi_j') = \min \left\{1,  \exp\left\{ -\theta (d_{oc}(\pi_j',\boldsymbol{z}) - d_{oc}(\pi_j^t,\boldsymbol{z})) \right\}   \right\}$ given current values of $\theta$ and $\boldsymbol{z}$. Iterating for $j = 1, \ldots, q$ gives the updated complete data $\underline{\pi}^{t+1}$ and $\boldsymbol{z}^{t+1} \sim p(\cdot|\underline{\pi}^{t+1}, \theta)$ and $\theta^{t+1} \sim p(\cdot|\underline{\pi}^{t+1}, \boldsymbol{z}^{t+1})$ are drawn as before.


\section{Inference: learning the CT structure}\label{sec:opt}

This section is devoted to optimizing the CMM probability of $\underline{\pi}$ for multiple CTs. Our goal here is to carry out fast optimization of $f(\underline{\pi}|\boldsymbol{z}, \theta)$ under a set of candidate structures. In other words, we would like to fit $\widehat{\boldsymbol{z}}$ and $\widehat{\theta}$ which are the maximum likelihood estimators of $\boldsymbol{z}$ and $\theta$. Once $\widehat{\boldsymbol{z}}$ and $\widehat{\theta}$ are obtained, model structure selection is based on the best possible allocation and spread under each given CT. The decision can be made with the information criterion in Equation (\ref{eq:info_criteria}) or a data-based metric that we will introduce later on.


Hence, fitting $\widehat{\boldsymbol{z}}$ and $\widehat{\theta}$ is pursued next with the main focus being on model selection. Our key concern here is to minimize the computational cost related to fitting the model for various CTs. Once a model structure is chosen, it can be used with the Bayesian method outlined in section \ref{sec:bayes}. This way, the posterior distributions of the chosen model parameters are derived, and uncertainty is adequately quantified. This type of two-step strategy is not novel, being well-known and effective in the Bayesian Networks (BNs) literature. In BNs, heuristic searches are commonly utilized to explore possible graph structures and choose the graph topology. Once a structure is selected, inferential methods like Markov Chain Monte Carlo are employed to estimate the model parameters of the chosen structure. For more details on Bayesian Networks inference see \cite{LARRANAGA2013109}, \cite{scutari2021bayesian},  \cite{koski2011bayesian}. Another related example is \cite{raftery2006}, where an information criteria-based strategy (BIC) is applied for a combined search of variable and cluster selection. 

The maximum likelihood estimators of $\boldsymbol{z}$ and $\theta$ are found as the solution of $\max_{\boldsymbol{z}, \theta} f(\underline{\pi}|\boldsymbol{z}, \theta)$. It is straightforward to see that $\widehat{\boldsymbol{z}}$ is the configuration that minimizes the sum of distances $\sum_{j=1}^q d_{oc}(\pi_j,\boldsymbol{z})$ given a CT. The $\boldsymbol{z}$ optimum does not depend on $\theta$ once $\min ( \sum_{j=1}^q d_{oc}(\pi_j,\boldsymbol{z}) ) = \min( \theta \sum_{j=1}^q d_{oc}(\pi_j,\boldsymbol{z}) )$ for any $\theta>0$. The MLE of $\theta$ follows conditionally on $\boldsymbol{z}$ by working out the solution to $S'(\theta) = 0$ where $S'(\theta)$ is the score equation for $\theta$. We start by outlining a procedure to derive $\widehat{\boldsymbol{z}}$ that uses simulated annealing. Given $\widehat{\theta}$, $\widehat{\theta}$ is estimated using the methodology of \cite{critchlow}.

\subsection{Allocations: Simulated Annealing Search}\label{sec:sim_anneal}

Simulated annealing (SA) is an optimization technique that relies on MCMC sampling. It can be straightforwardly applied to optimize $f(\underline{\pi}|\boldsymbol{z}, \theta)$ in terms of $\boldsymbol{z}$ using ideas introduced in previous sections. Maximizing the CMM probability with respect to the allocations $\boldsymbol{z}$ is equivalent to minimizing $\sum_{j=1}^q d_{oc}(\pi, \boldsymbol{z})$. Given that $\theta$ is greater than zero, the optimum $\boldsymbol{\widehat{z}}$ is the same for any $\theta$.

Denote by $g(\boldsymbol{z})$ the CMM probability as a function of $\boldsymbol{z}$ that takes observed data $\underline{\pi}$ and $\theta=1$. SA is based on the idea that powers $g(\boldsymbol{z})^\alpha$ make $f$ more concentrated around its mode as $\alpha \rightarrow \infty$. Metropolis (or MH) moves are used to sample from powers $\boldsymbol{z}^t \sim g(\cdot)^\alpha_t$ where $\alpha_t$ increases with iterations $t=1, \ldots, T$. The probability to move from $\boldsymbol{z}^t$ to $\boldsymbol{z}'$ is the Metropolis ratio $p(\boldsymbol{z}') = \min \{1, (g(\boldsymbol{z}')/g(\boldsymbol{z}^t))^{\alpha_t} \}$, under a symmetric proposal. As $\alpha_t$ increases, the transition probability converges to one if $g(\boldsymbol{z}')> g(\boldsymbol{z}^{t})$, zero otherwise. The sampler should eventually reach $\max_{\boldsymbol{z}} g(\boldsymbol{z})$ and stop accepting new moves. The key ingredient of SM is defining a discrete set of powers $\alpha_0 < \alpha_1< \cdots < \alpha_T$, which is called the \textit{annealing (or cooling) schedule}. It is often recommended to start from small $\alpha_t$ to avoid local optimums. 

An SA algorithm can be formulated in a similar spirit to the methods of section (\ref{sec:bayes}). As before, the $\texttt{switch}(\boldsymbol{z}^t, m)$ proposals generate perturbations $\boldsymbol{z}'$ and $p(\boldsymbol{z}')$ is

\begin{equation}
    p(\boldsymbol{z}') = \left( \frac{  \exp\left( - \sum_{j=1}^q d_{oc}(\pi_j, \boldsymbol{z}') \right)}{\exp\left( - \sum_{j=1}^q d_{oc}(\pi_j, \boldsymbol{z}^t) \right)} \right)^{\alpha_t}.
\end{equation}\label{eq:sa_prob}

It is easy to see that $\alpha_t$ plays the role of the CMM spread in the SA acceptance probability and the procedure is equivalent to the Metropolis algorithm of section (\ref{sec:bayes}). We can use this fact to guide the setup of $\boldsymbol{\alpha}$, the annealing schedule. The initial $\alpha_0$ should be close to 0 to allow for exploration of the $\boldsymbol{z}$ space. This starting temperature can then be increased by some fixed amount $s>0$ until a maximum $\alpha_T$. We recommend setting $\alpha_T$ from inspecting the CMM phase transition. This is the point where further increasing $\theta$ effectively causes no change in the distribution. Alternatively, the \textit{geometric schedule} is a popular choice. This is given by $\alpha_t = \beta^t \beta_0$ for some small $\beta_0$ and $\beta>1$.

An algorithmic description of this optimization routine if provided in the supplementary material (\ref{alg:cmm_sampler}). In what follows, it will be denoted by \textcolor{blue}{\textit{Annealing}($\underline{\pi}, \boldsymbol{\alpha}, m$)}, where $\boldsymbol{\alpha}$ is the annealing schedule and $m$ is the number of switches.


\subsection{$\theta$: Iterative Method}\label{sec:mle}

Given $\boldsymbol{\widehat{z}}$, the MLE of $\theta$ is found from the solution of $S'(\theta)=0$. This is given by \newline $-\theta \sum_{j=1}^q d_{oc}(\pi_j,\boldsymbol{z}) \exp \{ -\theta \sum_{j=1}^q d_{oc}(\pi_j,\boldsymbol{z}) \} \Psi(\theta)^{-q} -q  \Psi(\theta)^{-q-1}   \exp \{ -\theta \sum_{j=1}^q d_{oc}(\pi_j,\boldsymbol{z}) \}  =0$. It is shown in \cite{critchlow} that, with some algebraic manipulation, we can write this as

\begin{equation}\label{eq:theta_mle}
 E_{f} \left[ d_{oc}(\pi;\widehat{\boldsymbol{z}}) \right] =     \frac{1}{q} \sum_{j=1}^q d_{oc}(\pi_j, \widehat{\boldsymbol{z}}),
\end{equation}
an expectation with respect to $f(\cdot|\theta,\widehat{\boldsymbol{z}})$. The implication of (\ref{eq:theta_mle}) is that if $E_{f}$ can be estimated it is possible to optimize $\theta$ in a method of moments fashion. This means that $\widehat{\theta}$ is such that 
$E_{f} \left[ d_{oc}(\pi,\widehat{\boldsymbol{z}}) \right]$ is as close as possible to $\bar{d} = \sum_{j=1}^q d_{oc}(\pi_j, \widehat{\boldsymbol{z}})/q$ the empirical average distance. The proof of Equation (\ref{eq:theta_mle}) for the CMM can be found in the supplementary material. Using this result, we propose an iterative search that estimates $E_{f}$ with Monte Carlo. 

Algorithm (\ref{alg:mle}) describes this procedure in a summarisation of CMM maximum likelihood estimation. In STEP 1, $\widehat{\boldsymbol{z}}$ is fitted with SA, and the empirical $\bar{d}$ is computed from the data. The estimation of $\theta$ takes an initial guess $\theta^0$ and a pre-specified tolerance $\epsilon>0$. Auxiliary data is sampled in STEP2 using the current $\theta^t$ and Algorithm (\ref{alg:cmm_sampler}). The number of samples is $\lceil 1/\epsilon \rceil$, inversely proportional to the tolerance. In STEP 3, the expectation in (\ref{eq:theta_mle}) is estimated by the Monte Carlo average $\bar{d}_{\theta^t} = (\sum_{j=1}^{\lceil 1/\epsilon \rceil} d_{oc}(\pi_j, \boldsymbol{\widehat{z}}))/\lceil 1/\epsilon \rceil$, where $\pi_j' \approx CMM(\boldsymbol{\widehat{z}}, \theta^t)$. STEP 4 adjusts $\theta^t$ according to the relative difference between $\bar{d}$ and $\bar{d}_{\theta^t}$, setting $\theta^{t+1} \leftarrow \theta^t(\bar{d}/\bar{d}_{\theta^t})$. The algorithm stops once the absolute difference $|\theta^{t+1} - \theta^t|$ is below $\epsilon$.

\begin{algorithm}[]
\small
\KwIn{${\underline{\pi}}$, $N$: Gibbs sampler iterations, $\boldsymbol{\alpha}$: annealing schedule, CT, $\theta^0$: initial value of $\theta$, $\epsilon>0$: tolerance. }

\textcolor{gray}{STEP 1:} Obtain $\widehat{\boldsymbol{z}} \leftarrow \textcolor{blue}{\mbox{\textit{Annealing}}(\underline{\pi}, \boldsymbol{\alpha}, m)}$;

\hspace{0.8cm} Compute $\bar{d} = \sum_{j=1}^q d_{oc}(\pi_j, \widehat{\boldsymbol{z}})$;

Set $C = 0$; \textcolor{Cerulean}{// Convergence indicator}

$\theta^t \leftarrow \theta^0$;  \textcolor{Cerulean}{// Initialise $\theta$}

\While{$C = 0$}{

\textcolor{gray}{STEP 2:} $\underline{\pi}' \leftarrow  \textcolor{blue}{rCMM(\widehat{\boldsymbol{z}}, \theta^t, N, \lceil 1/\epsilon \rceil)}$; \textcolor{Cerulean}{// Sample the CMM$(\widehat{\boldsymbol{z}}, \theta^t)$  using Algorithm (\ref{alg:cmm_sampler})}

\textcolor{gray}{STEP 2:}  Compute $\bar{d}_{\theta^t} =\sum_{j=1}^{\lceil 1/\epsilon \rceil} \frac{d_{oc}(\pi_j',\boldsymbol{\widehat{z}})}{\lceil 1/\epsilon \rceil}$; \textcolor{Cerulean}{// Distance expectation from Monte Carlo}

\textcolor{gray}{STEP 4:} Set $\theta^{t+1} \leftarrow \theta^t \left ( \dfrac{{\bar{d}}_{\theta^t}}{\bar{d} } \right)$;

\uIf{$ |\theta^{t+1} - \theta^{t}| < \epsilon$ }{
    $C = 1;$
} \Else{
$\theta^t \leftarrow \theta^{t+1}$
}

}

\textbf{Output:} $\widehat{\theta} = \theta^{t}$ and $\boldsymbol{\widehat{z}}$;

\caption{\textcolor{blue}{$optCMM(\underline{\pi}, \boldsymbol{\alpha}, \epsilon, N)$} Optimization of the CMM model parameters $\boldsymbol{z}$ and $\theta$ with observed data $\underline{\pi}$.}\label{alg:mle}
\end{algorithm}

Algorithm \textcolor{blue}{$optCMM(\underline{\pi}, \boldsymbol{\alpha}, \epsilon, N)$} outputs $\{\widehat{\boldsymbol{z}}, \widehat{\theta} \}$ at low computational cost and is exploited next for CT selection. We conclude with the remark that Algorithm (\ref{alg:mle}) also applies to fitting the CMM in the frequentist approach. While uncertainty quantification of $\{\widehat{\boldsymbol{z}}, \widehat{\theta} \}$ is not readily, it is feasible to obtain with bootstrapping techniques. This involves optimizing re-sampled or replicated data $\underline{\pi}^r$ for $r=1, \ldots, R$ with the same algorithm.


\subsection{CT search}\label{sec:greedy_search}

Using estimates from Algorithm (\ref{alg:mle}) and a selection criterion, we can design an algorithmic search to find the CT that best fits the observed data. Our suggestion is to carry out a greedy exploration of the CT space, which aims to find the optimal model by proposing modifications of an initial $CT^t$. A greedy method finds a local optimum by moving in the direction that improves the criterion at each step, stopping once no improvement is achieved. To this end, this subsection addresses two key points: supplying a sensible initial value and defining a selection criterion. 

To provide a sensible initial guess, we propose to find a starting $CT$ that is guided by the preference proportions matrix described in section \ref{sec:intro}. This involves computing the empirical proportions $\bar{p}(i, i')= \sum_{j=1}^{q} I\{ \pi_j(i) \succ \pi_j(i') \}/n$ for all $i \neq i'$, $i, i' \in \{1, \dots, n\}$. This matrix can be used to construct an initial CT by observing that pairs with $\bar{p}(i, i')$ close to 0.5 are near indifference. Hence, an iterative scan of $\bar{p}(i, i')$ can be performed to group elements that are close to 0.5 within a certain tolerance. For example, if this is $tol = 0.05$, pairs such that $\bar{p}(i, i')$ is within (0.45, 0.55) are candidates to be grouped. In our strategy, if there is more than one $\{i, i'\}$ combination that satisfies this criterion, the one with the minimal absolute difference to 0.5 is chosen first. The search continues until there are no $|\bar{p}(i, i') -0.5|<tol$ that have not been grouped. As this search evolves and items are placed together, $\bar{p}(\cdot, \cdot)$ involving at least one group takes all possible element-wise combinations. For example, the starting CT table for the sushi problem with $tol = 0.05$ takes four iterations. In the first step (top-left window), \texttt{squid} and \texttt{tuna roll} are grouped, as this is the pair with $\bar{p}$ closest to 0.5 ($\bar{p}(\texttt{squid}, \texttt{tuna roll}) = 0.51$). In step two, the preferences proportion matrix is recomputed, and this is now a $(9 \times 9)$ matrix where $\bar{p}( \{\texttt{squid}, \texttt{tuna roll}\}, i')$ is $\frac{1}{2\times 280} (\sum_{j=1}^{280} I\{ \pi^{-1}_j(\texttt{squid}) <  \pi^{-1}(i') \} + I\{ \pi^{-1}_j(\texttt{tuna roll}) <  \pi^{-1}(i') \}$ when comparing this group to some sushi other $i'$. 

\begin{figure}
    \centering
    \includegraphics[width =0.4\linewidth]{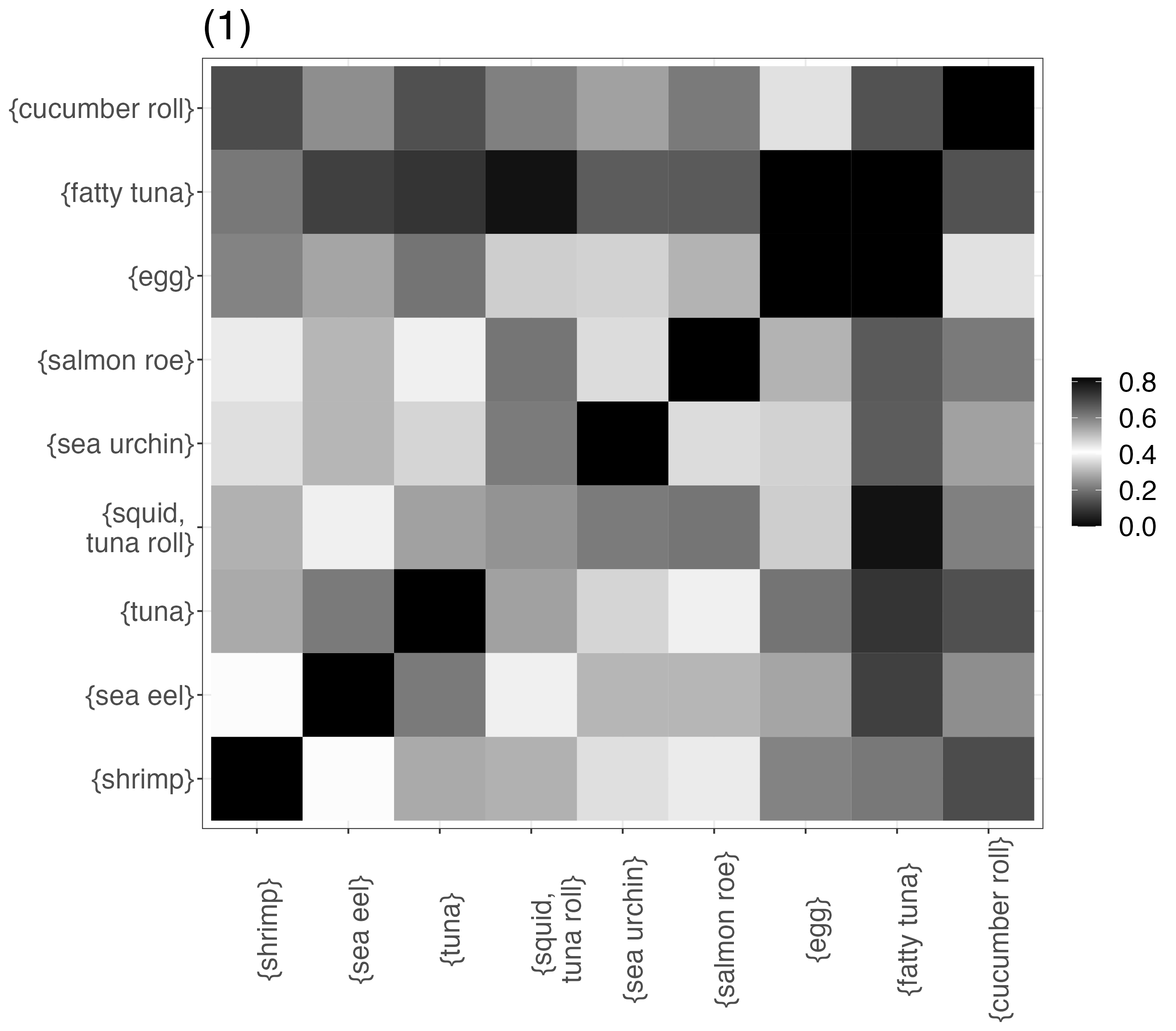}
    \includegraphics[width =0.4\linewidth]{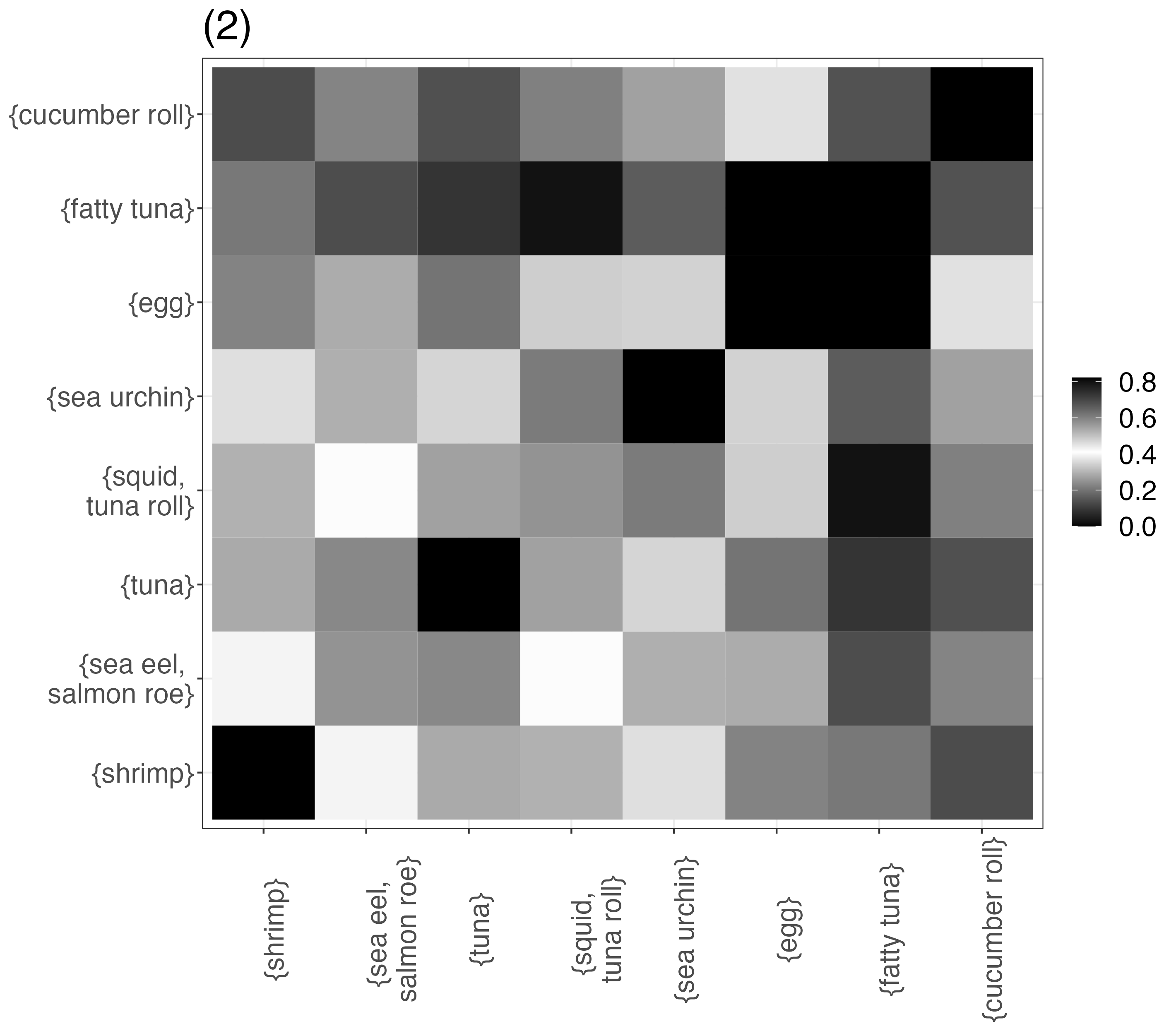}
    \includegraphics[width =0.4\linewidth]{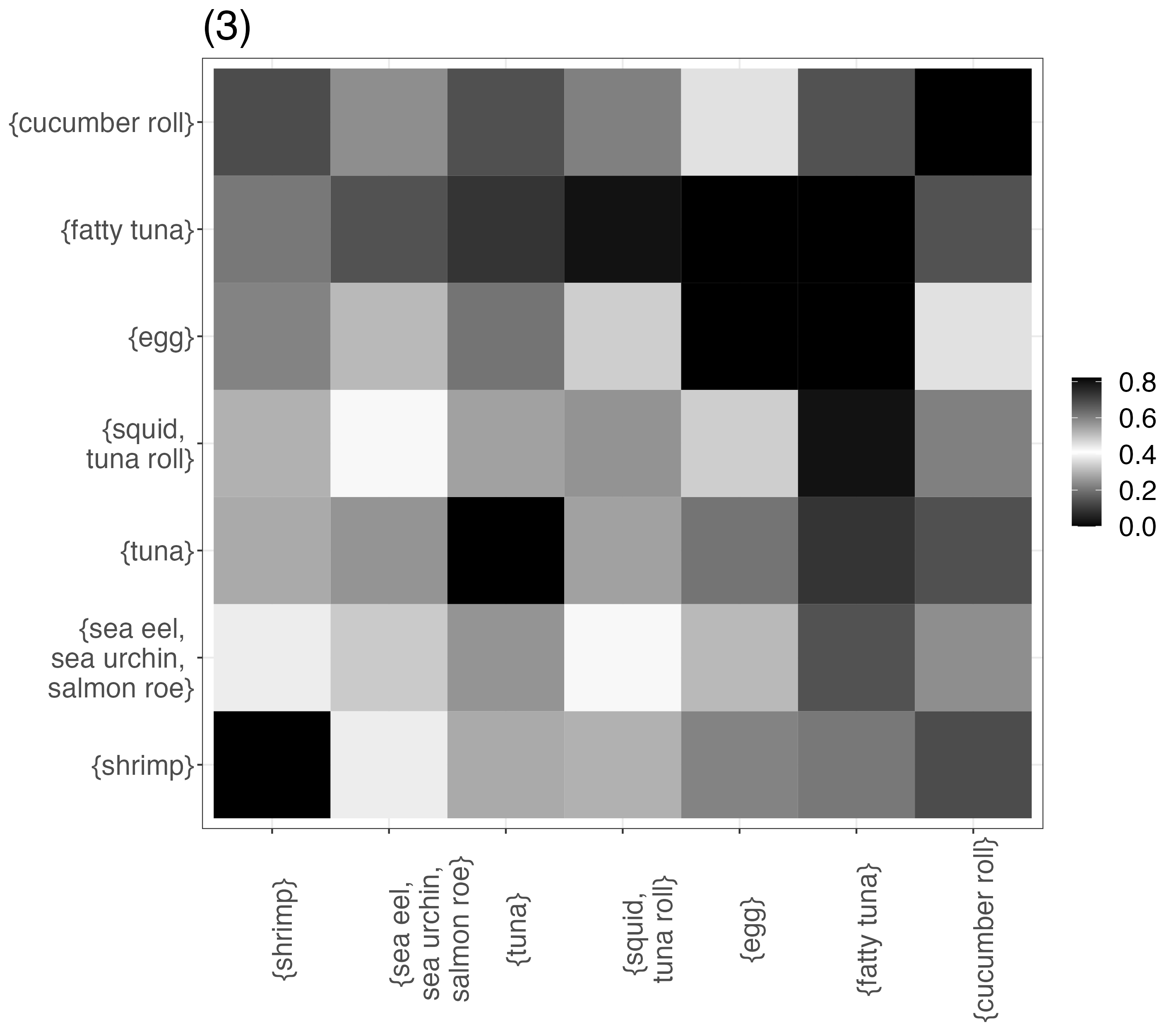}
     \includegraphics[width =0.4\linewidth]{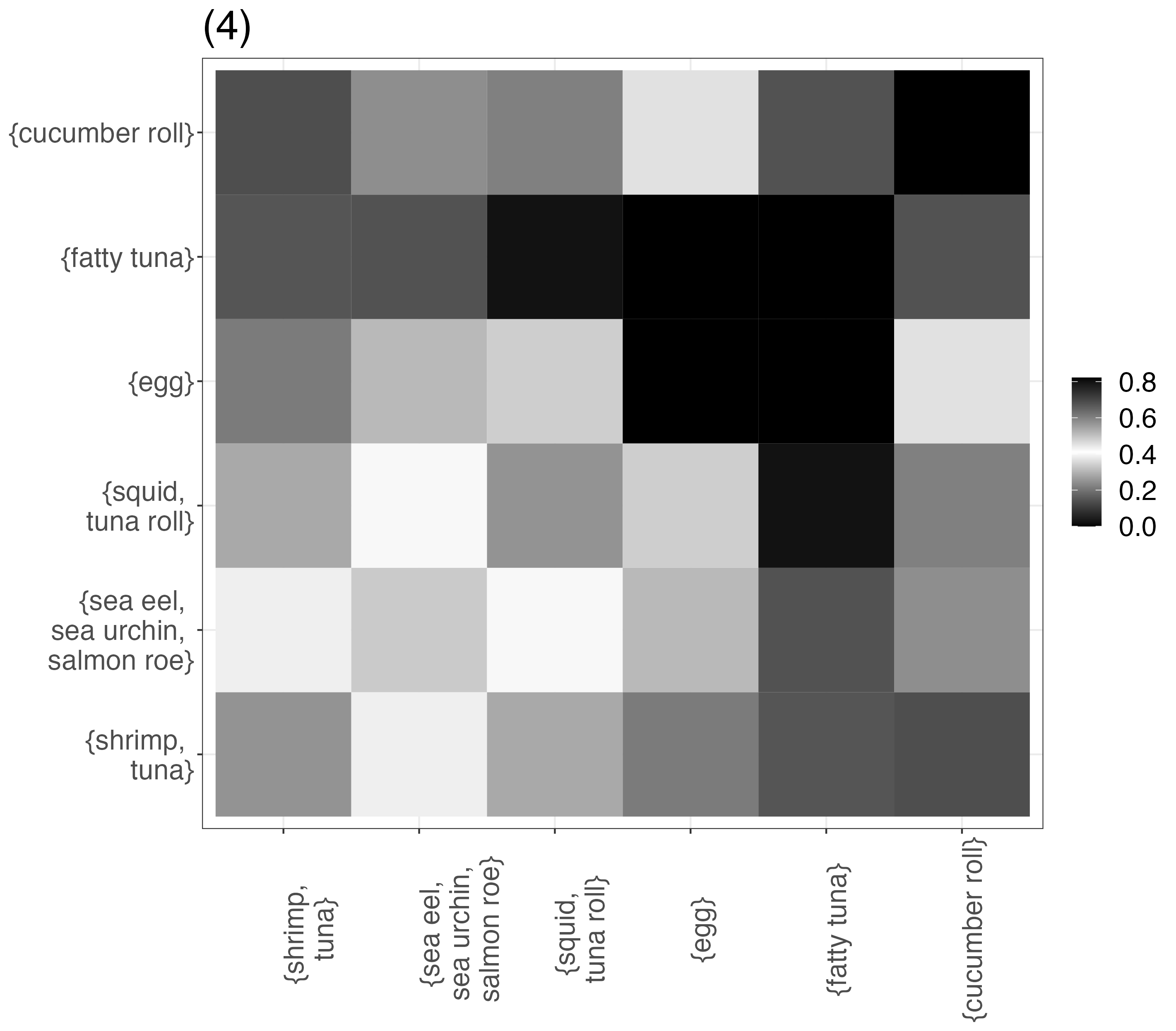}
    \caption{Aggregation of the sushi's pairwise preferences matrix based on the closeness between $\bar{p}(i, i')$ to indifference. Items (or clusters) are merged iteratively by checking if the average empirical preferences are within $tol =0.05$ to 0.5. The algorithm takes four iterations which are indicated at the top of each plot. }\label{fig:sushis_init_CT}
\end{figure}

Aggregation of the sushis pairwise preferences matrix based on the described procedure is illustrated in Figure \ref{fig:sushis_init_CT}. The steps taken to arrive at the initial $CT$ for this data are organized as rows, with their corresponding iteration number shown at the top. The final group sizes are (1, 1, 1, 2, 2, 3), as indicated on the bottom-right window. We recommend $tol$ close to $\sqrt{0.5/q}$ as a canonical choice of this parameter. This formulation takes into account that the standard deviation of a proportion $\widehat{p}$ estimated from $q$ samples depends on $q$. If $\widehat{p}$ is near 0.5, we can set $tol \approx sd(\widehat{p}) = \sqrt{ \widehat{p}(1/q) }$. In the previous example, $\sqrt{0.5/280} = 0.042$.

Once an initial CT is obtained, modifications of it are proposed and evaluated in terms of a selection criterion. Denoted by $\mathcal{N}(CT^t)$, these are \textbf{neighboring tables} of the current $CT^t$ that can be designed in different ways. For example, a \textit{one adjacent shift} builds $\mathcal{N}(CT^t)$ by shifting one observation from cluster $l$ to $(l-1)$ or $(l+1)$. All those possible from $CT^t = (1,3,3,3)$ are $\mathcal{N}(CT^t) = \{ (4,3,3), (2,3,3,3), (1,2,4,3), (1,4,2,3), (1,3,2,4), (1,3,4,2), \newline (1,3,3,2,1) \}$. The number of tables can be up to $2\times L$ so an exhaustive search under \textit{one adjacent shifts} is not feasible for large $L$. In this case, a possible mitigation is to explore sampled subsets of the possible $\mathcal{N}(CT^t)$. Changing the type of search moves and setting restrictions on the $\boldsymbol{z}$ space are other possible strategies. While adjacent shift searches simultaneously $L$ and $n_1, \ldots, n_L$, one could fix $L$ or some $n_l$. For example, fixing $n_l=1$ for low and high $l$ explores the hypothesis of \textit{top-bottom elicitation}. Once $\mathcal{N}(CT^t)$ is determined, models are optimized with the routine \texttt{optCMM()} and the selection criterion is computed. For computational efficiency, visited configurations and their criterion values can be stored and recycled across the search. 

The natural selection criterion to employ in the CT search is (\ref{eq:info_criteria}). Analogously to AIC and BIC methods, this criterion works with the optimized log-likelihood value and employs a penalisation for model complexity. However, to compute (\ref{eq:info_criteria}), it is necessary to estimate the CMM likelihood with importance sampling. In addition, the CT search requires doing so with enough precision so that Monte Carlo variability does not impact the algorithm's decisions. Evidently, the variability to which $\hat{\mathcal{I}}(CT)$ is estimated can be made smaller by increasing $M$ but the computational cost increases. To tackle this situation, we stipulate and test an alternative criterion that is purely data-based. 

A data-based goodness of fit metric can be formulated on top of the data's agreement to the proposed partitions. Once a $CT$ is proposed and $\hat{\boldsymbol{z}}$ is found, we can compute the proportion of times that rankings in $\underline{\pi}$ agree with the preference relations in $\hat{\boldsymbol{z}}$. This is computed similarly to what is done in the initial value procedure but does not involve a pre-specified tolerance. For example, with the ordered partition $\{ \boldsymbol{z} \}$ = (\{1,2\},\{ 3,4\},\{ 5\}) there are 3 unique clusters, making $3 \choose 2$ between-group comparisons necessary. Comparisons are made across all between-cluster element combinations, checking how often the smaller-label object precedes the higher-label one in the data. In the example, clusters (1,2), (1,3) and (2,3) are compared. For the first pair, this consists of evaluating the between-set combinations of $\{ 1,2\}$ and $\{ 3,4\}$. Hence, the contrast of clusters one and two are done with $\sum_{j=1}^q I\{ \pi^{-1}_j(1) < \pi^{-1}_j(3) \} +$ $\sum_{j=1}^q I\{ \pi^{-1}_j(2) < \pi^{-1}_j(3) \} +$ $\sum_{j=1}^q I\{ \pi^{-1}_j(1) < \pi^{-1}_j(4) \} +$ $\sum_{j=1}^q I\{ \pi^{-1}_j(2) < \pi^{-1}_j(4) \}$. After this is computed for all between-group combinations, averaging across the total number of comparisons, i.e. $q \times \left({ n \choose 2} - \sum_{l=1}^L {n_l \choose 2}  \right)$, renders the proportion of correct preference relations. Similarly to with $\widehat{\mathcal{I}}(CT)$, the goal is to maximize this proportion. This metric is denoted by $\mathcal{D}(CT, \underline{\pi})$ and satisfies $0<\mathcal{D}(CT, \underline{\pi})<1$ where values close to one are indicative of a better fit.

\begin{algorithm}[]
\small
\KwIn{$CT^t$: initial CT, ${\underline{\pi}}$: observed data, $\boldsymbol{\alpha}$: annealing schedule, $\epsilon$: optimization tolerance for $\theta$, $N$: Gibbs sampler iterations, $M$: IS replicas;  }

\vspace{0.2cm}
Optimize $CT^t$: $\{ \boldsymbol{\widehat{z}}, \widehat{\theta} \}^t \leftarrow$ \textcolor{blue}{$optCMM(CT^t, \underline{\pi}, \boldsymbol{\alpha}, \epsilon, N)$};

Estimate $\widehat{\mathcal{I}}(CT^t)$ with $\{ \boldsymbol{\widehat{z}}, \widehat{\theta} \}^t$;

\vspace{0.2cm}

$\mathcal{L} = ( CT^t,  \widehat{\mathcal{I}}(CT^t)$; \textcolor{Cerulean}{// Initialise list of visited configurations}

\vspace{0.2cm}

\textcolor{gray}{STEP 1:} List neighboring tables: $\mathcal{N}(CT^t) \equiv \{ CT^t_1, \ldots, CT^t_J\}$;  \textcolor{Cerulean}{// $CT^t$ modifications}

\vspace{0.2cm}
 
\textcolor{gray}{STEP 2:} \For{$j \in 1:J$}{

\uIf{ $CT^t_j \notin \mathcal{L}$ \textcolor{blue}{// Table not visited} }{ 
    $\{ \boldsymbol{\widehat{z}}, \widehat{\theta} \}^t_j \leftarrow$ \textcolor{blue}{$optCMM(CT^t_j, \underline{\pi}, \boldsymbol{\alpha}, \epsilon, N)$};

Compute $\widehat{\mathcal{I}}(CT^t_j)$ and append  $CT^t_j, \widehat{\mathcal{I}}(CT^t_j)$ to $\mathcal{L}$;
} \Else{ 

Recycle $\widehat{\mathcal{I}}(CT^t_j)$ from $\mathcal{L}$; \textcolor{Cerulean}{// Table visited}
}
}

\textcolor{gray}{STEP 3:} \uIf{ $\max\{  \widehat{\mathcal{I}}(CT^t_1), \ldots, \widehat{\mathcal{I}}(CT^t_J) \} > \widehat{\mathcal{I}}(CT^t)  $}{

Update $CT^t$ to the $CT^t_j$ with maximum $\widehat{\mathcal{I}}(CT^t_j)$;

Return to \textcolor{gray}{STEP 1};
} \Else{
\textcolor{gray}{STOP};
}

\textbf{Output:} $CT^t$;

\caption{Greedy search of the CMM's CT using $\widehat{I}(CT)$. Selection with the data-based criterion $\mathcal{D}(CT, \underline{\pi})$ is done similarly as does not require sampling in STEP 2.}\label{alg:search}
\end{algorithm}

Algorithm (\ref{alg:search}) summarises with pseudo-code the CMM CT search. The algorithm is written in terms of $\hat{\mathcal{I}}(CT)$ but works equivalently with $\mathcal{D}(CT, \underline{\pi})$, the data-based criterion. In STEP 1, modifications of $CT^t$ are proposed, and this is the neighboring tables set, $\mathcal{N}(CT^t)$. STEP 2 uses the routine \texttt{optCMM()} to find the best ${\boldsymbol{z}}$ and $\theta$ for each model structure in $\mathcal{N}(CT^t)$ and the optimized model parameters are used to compute $\hat{\mathcal{I}}(CT)$ (or $\mathcal{D}(CT, \underline{\pi})$). If the proposed $CT$ has already been visited, STEP 2 is recycled from previous iterations. In STEP 3, the algorithm moves in the direction that optimizes the criterion, stopping if no improvement is achieved.

\subsubsection{Simulation studies}

\begin{figure}
    \centering
    \includegraphics[width = 0.6\linewidth]{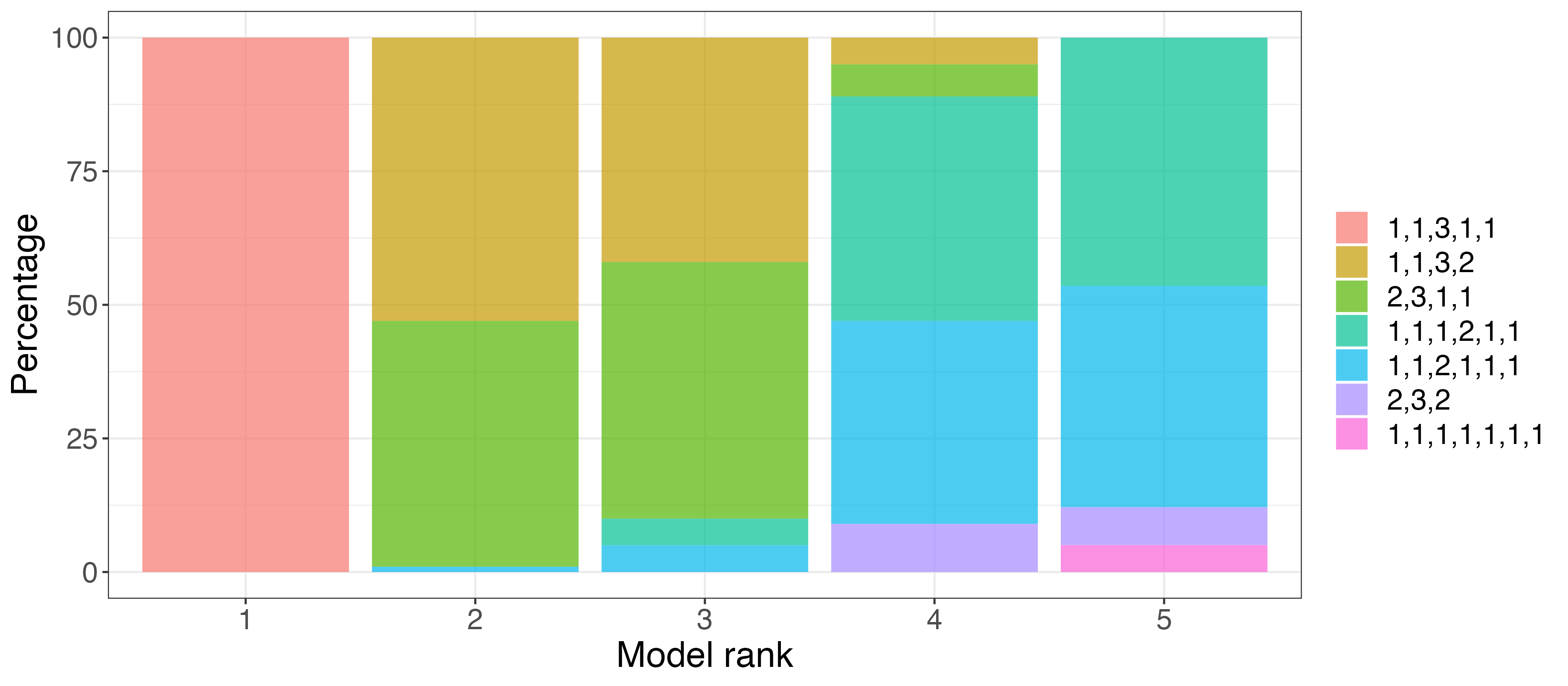}
    \caption{ Summary of top five models from 100 searches of $n(\boldsymbol{z}) = (1,1,3,1,1)$. Each bar indicates the proportion of times a configuration appeared as the first to fifth choice.}
    \label{fig:simulation_1}
\end{figure}

In what follows, we evaluate the CT search with simulation studies. Three setting are explored to investigate different aspects of the proposed methodology. In the first, one hundred CMM data sets of $q=500$ ranks are simulated using the Hamming ordered cluster distance. The CT and $\theta$ used to generate the data are $(1,1,3,1,1)$ and $\theta = 0.9$. This configuration is designed to inspect the CT search with the optimization of $\widehat{\mathcal{I}}(\cdot)$. Under small $n=7$, we set $M = 100K$ and run Algorithm (\ref{alg:search}) with one-adjacent shifts in STEP 1.  The search is done for each for each replicated data $\underline{\pi}^r$, $r\in 1:100$ and the CTs with the five highest $\mathcal{I}$ values are stored. The results are summarised in Figure \ref{fig:simulation_1} where the horizontal axis shows the model rank (first to fifth) and the bars are colored according to the proportion of times a model was ranked in that position. We can see in the first (from left to right) bar that the CT used to generate the data is selected every time as the best candidate. Inspection of the following ranks is useful to show that the next-to-best ones are in close resemblance with $(1,1,3,1,1)$. For instance, choices in the second and third to best places differ from the true CT by one adjacent shift.

\begin{figure}
    \centering
    \includegraphics[width = 0.8\linewidth]{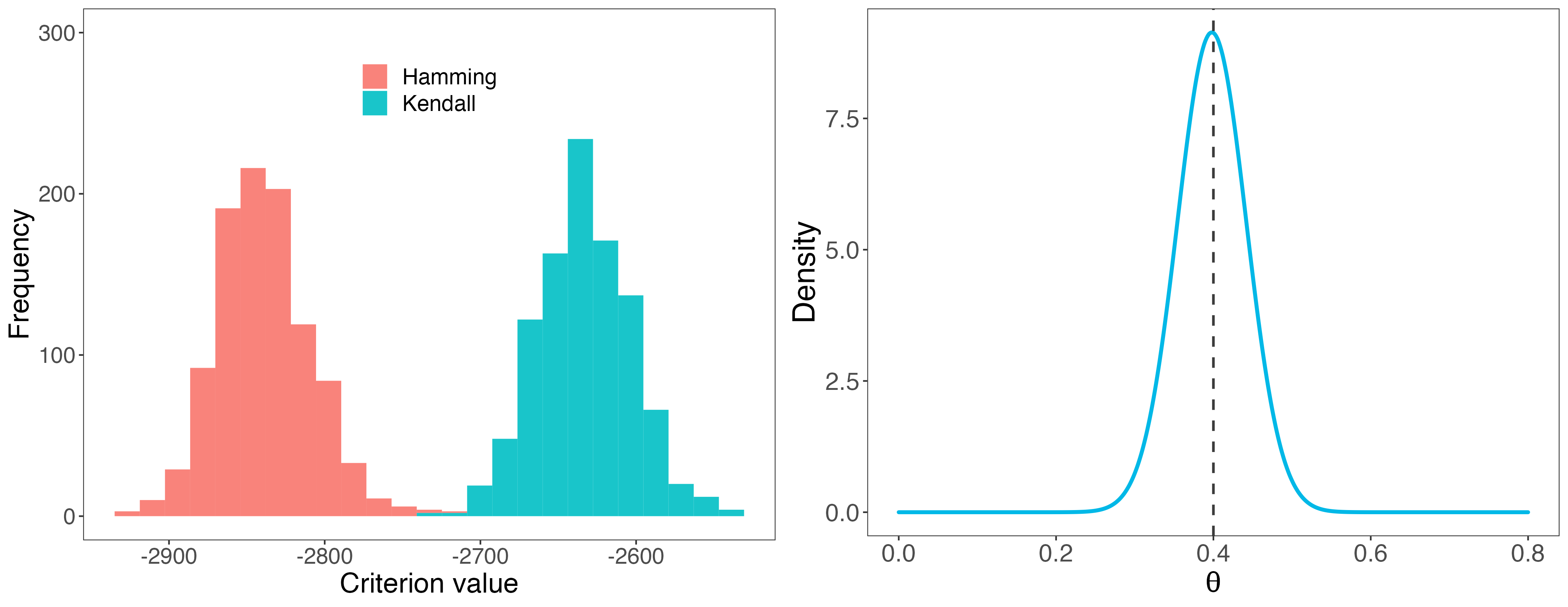}
    \caption{To the left, histograms of $\mathcal{I}$ values from 1K fits of the Kendall and Hamming CMM to CMM-k simulated data. Estimates of the spread $\theta =0.4$ from the correctly-specified model are shown to the right. The value of $\theta$ used to generate the data is indicated by the vertical dashed line.}
    \label{fig:dist_criterion_and_theta}
\end{figure}

A second study is concerned with the choice of distance. In this setting, we use the Kendall $d_{oc}$ to draw $\underline{\pi}^r \sim CMM(\theta, \boldsymbol{z})$ with $q = 200, n=10$ for $r = 1,\ldots, 1K$. The parameters used to simulate the data are $\boldsymbol{z} = (1,2,2,3,3,3,3,4,4,5)$ and $\theta = 0.4$, so $CT = (1,2,4,2,1)$. The CMM is fitted to each $\underline{\pi}^r$ using the Hamming and Kendall ordered cluster distances under the correctly specified CT. Computing $\mathcal{I}(\cdot)$, we obtain 1K estimates of the criterion from $d_{og,k}$ and $d_{og,h}$ model fits. These are displayed with histograms in Figure \ref{fig:dist_criterion_and_theta}, to the left, showing that the criterion selects the correct distance, as desired. The values of $\mathcal{I}$ obtained with the Kendall ordered cluster distance are higher and share negligible overlap with those from the Hamming fits.

We also stored $\{ \boldsymbol{\widehat{z}}, \widehat{\theta}\}$ from this study to investigate the performance of Algorithm (\ref{alg:mle}). Ideally, we would like the correctly-specified model to produce $\widehat{\theta}$ estimates close to $\theta = 0.4$ which was used to simulate the data. Figure \ref{fig:dist_criterion_and_theta} shows that this is achieved, indicating the true value with a vertical dashed line. Fitted values of $\widehat{\boldsymbol{z}}$ are also explored and the correct configuration is recovered about 95\% times with either $d_{oc}$. The complete table of frequencies is reported in the supplement. The analysis of $\boldsymbol{\widehat{z}}$ allows us to conclude that its estimation is robust with respect to the distance. In other words, cluster allocations are well estimated under misspecification of $d_{oc}$ which is a desirable feature.

\begin{figure}
    \centering
    \includegraphics[width =0.4\linewidth]{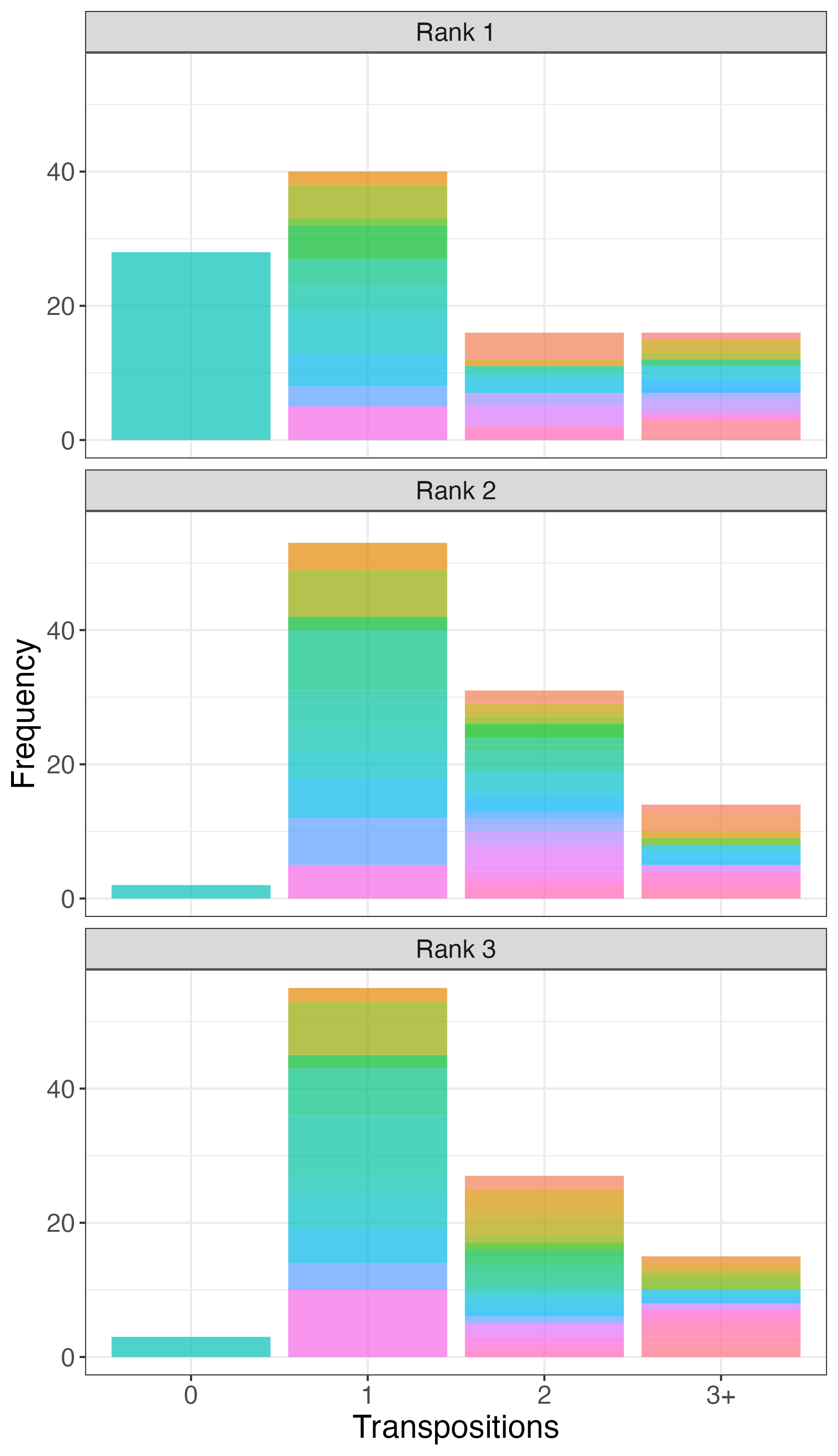}
    \includegraphics[width =0.4\linewidth]{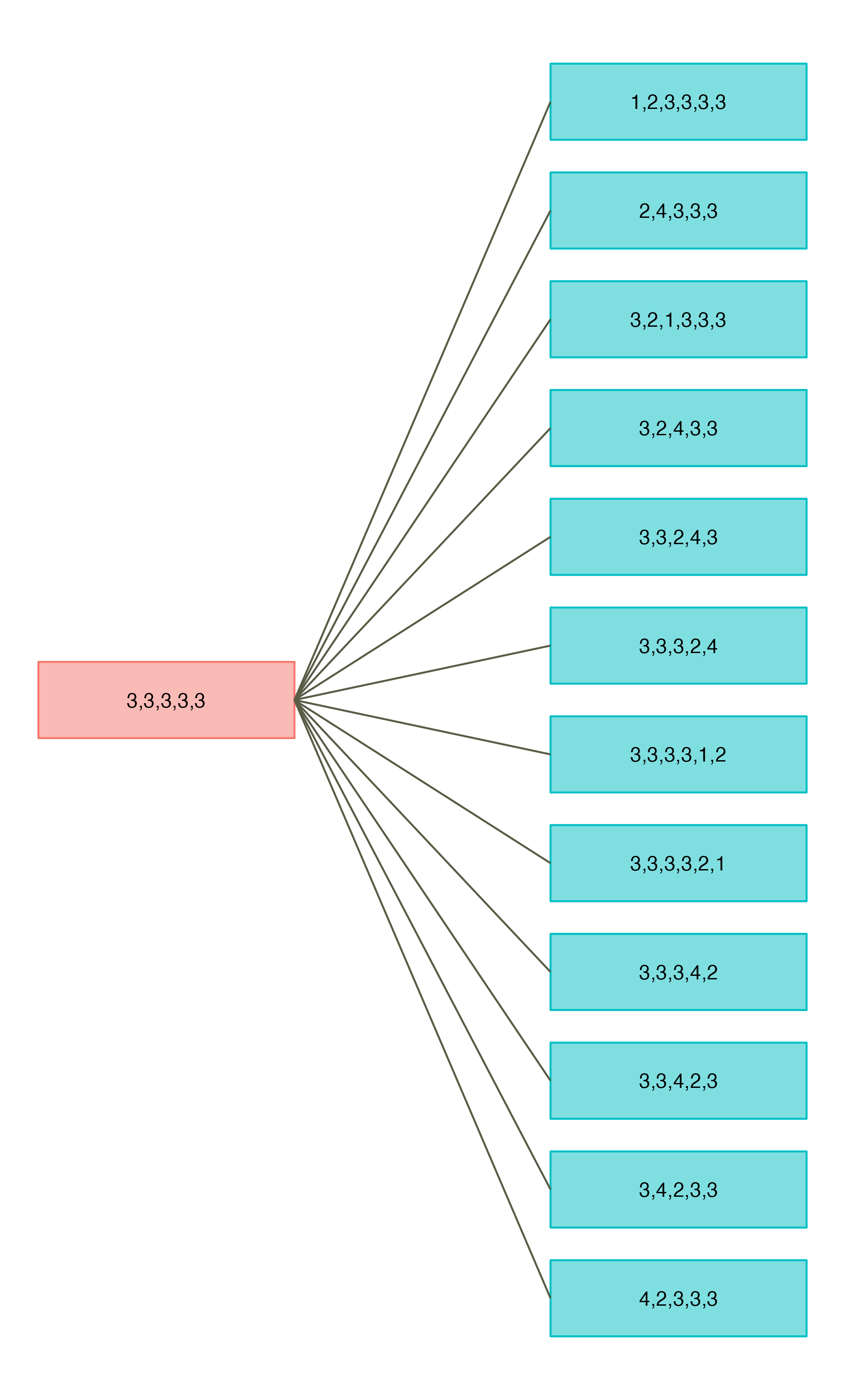}

    \caption{To the left, frequency of CTs selected with Algorithm (\ref{alg:search}) as best (top panel), second-to-best (middle panel) and third-to-best (bottom panel). Within each window, bars show the aggregation of frequencies for CTs within zero, one, two or three or more single-adjacent-shifts from the true CT. On the right, a diagram exemplifies different tables that are all within one single-adjacent-shift from (3,3,3,3,3). Categories shown in blue compose the second bar in the top window plot, where they are indicated in different colors. }\label{fig:sim_experiment3}
\end{figure}

Our final simulation study illustrates the CT search under a moderate set of $n=15$ items and $\mathcal{D}(\cdot, \underline{\pi})$. In this experiment, the CT used to simulate the data is $CT = (3,3,3,3,3)$ and $\theta = 1.5$. One hundred replications are considered as before, and we summarise the results in Figure \ref{fig:sim_experiment3}. To the left, three panels are included to illustrate the frequency of the $CT$s chosen as best (first row), second-to-best (second row) and third-to-best (last row). In each panel, the four bars aggregate the frequency of CTs that are within zero, one, two or three or more adjacency shifts from the true table. Within each bar, colors are displayed to indicate possible CTs in that distance from (3,3,3,3). Naturally, only the correct table is within zero shifts from the true CT, so the first bar is the frequency of exactly right matches. A large frequency of one-shift results is seen in the three plots which may at first seem to surpass the correct CT frequency. However, we point out that this is composed by a large number of tables such as the ones on the diagram to the right. It shows the CTs that are selected (ranked first) by the search and can be obtained from the true CT with a single adjacent shift. With this, we bring attention to the fact that CTs within very close resemblance to the generating one are often selected. We can conclude that although the search based on $\widehat{\mathcal{I}}(\cdot)$ provides more accurate results, $\mathcal{D}(\cdot, \underline{\pi})$-based selection provides a close approximation. When the cost of estimating $\widehat{\mathcal{I}}(\cdot)$ with small variability is prohibitive, the data-based selection is a suitable alternative. The two methods can also be employed in conjunction by using $\mathcal{D}(\cdot, \underline{\pi})$ to narrow the search and then compute $\widehat{\mathcal{I}}(\cdot)$ for a smaller subset.

\section{Data applications}\label{sec:applications}

\subsection{Formula 1 data analysis}\label{sec:formula1}

In this section, we analyse the final rank of drivers in races of the Formula 1 drivers championship. This data is available online from a variety of sources, and we focus on the races which took place in 2015 and 2016. This a set of $q=40$ races where 28 different drivers participated at least once. Each circuit starts with 20 cars so rankings of up to 20 competitors are recorded. There can be a smaller number of finishers if drivers fail to finish the race. We choose to filter out competitors that participated 10 times or less, resulting in an item set of $n=25$ for analysis. The data is then viewed as an incomplete ranking in $\mathcal{P}_{25}$ where races $j =1, \ldots, 40$ are coded by $\widetilde{\pi}_j$. This is a partial ranking observation of the $n=25$ drivers set.

Our goal is to analyse $\underline{\widetilde{\pi}} \equiv (\widetilde{\pi}_1, \ldots, \widetilde{\pi}_{40})$ with the CMM and a CT guided by the Formula 1 point system. The competition awards drivers with scores according to their rank in each race. The championship winner is then the driver with the most points at the end of the season.
There are some subtle variations, but the basic scoring rule is that points are given to the top-10 finishers as (25,18,15,12,10,8,6,4,2,1), for the first to tenth, respectively.

Based on this, we set the CMM CT to $(1,1,1,2,3,2,15)$. Podium positions that are worth 25, 18, and 15 are each their own cluster, i.e., $n_1 = n_2 = n_3 = 1$. This is followed by grouping the fourth and fifth places awarded 10 and 12 points so $n_4 = 2$. Scores \{4,6,8\}, \{1,2\} and \{0\} are the last clusters, making $n_4 = 3, n_5 = 2$ and $n_6 = 15$. The idea behind this partition choice is to allow interchangeability between drivers within the above scoring ranges. For example, it is reasonable to assume that there are competitors who most often do not score (cluster 6). 

The unclustered MM is also fitted to $\underline{\pi}$, where drivers are strictly better or worse than one another in the consensus $\pi_0$. Both models are fitted under the Bayesian approach and samples from the posterior distributions of $(\theta, \boldsymbol{z})$ and $(\pi_0, \alpha)$ are collected. The Mallows model is fitted with the R package \texttt{BayesMallows} \citep{bayesmallows} that implements an equivalent augmentation strategy.

Posterior modes of $\boldsymbol{z}$ and $\pi_0$ are computed after running each algorithm for 11K iterations and discarding the first 1K as burn-in. The top 10 drivers in $\boldsymbol{z}$ and $\pi_0$ are summarised in Table \ref{tab:modes_drivers}. The CMM and MM show agreement between the model fits and their difference is that some strict ordering is estimated in the latter for the clustered drivers.

\begin{table}[H]
\centering
\begin{tabular}{@{}ccc@{}}
\toprule
\textbf{Points} & \textbf{CMM}                    & \textbf{MM}                            \\ \midrule
25              & \{Hamilton\}                  & Hamilton                               \\
18              & \{Rosberg\}                   & Rosberg                                \\
15              & \{Vettel\}                    & Vettel                                 \\
\{12, 10\}      & \{Raikkonen, Ricciardo\}      & Raikkonen $\succ$ Ricciardo            \\
\{ 8,6,4 \}     & \{Bottas, Massa, Verstappen\} & Bottas $\succ$ Massa $\succ$Verstappen \\
\{ 2,1\}        & \{Grosjean, Perez\}            & Perez $\succ$ Grosjean                 \\ \bottomrule
\end{tabular}
\caption{Posterior mode of CMM ordered clusters $\{ \boldsymbol{z}\}$ and MM's central permutation $\pi_0$. }\label{tab:modes_drivers}
\end{table}

\begin{figure}
    \centering
    \includegraphics[width = 0.8\linewidth]{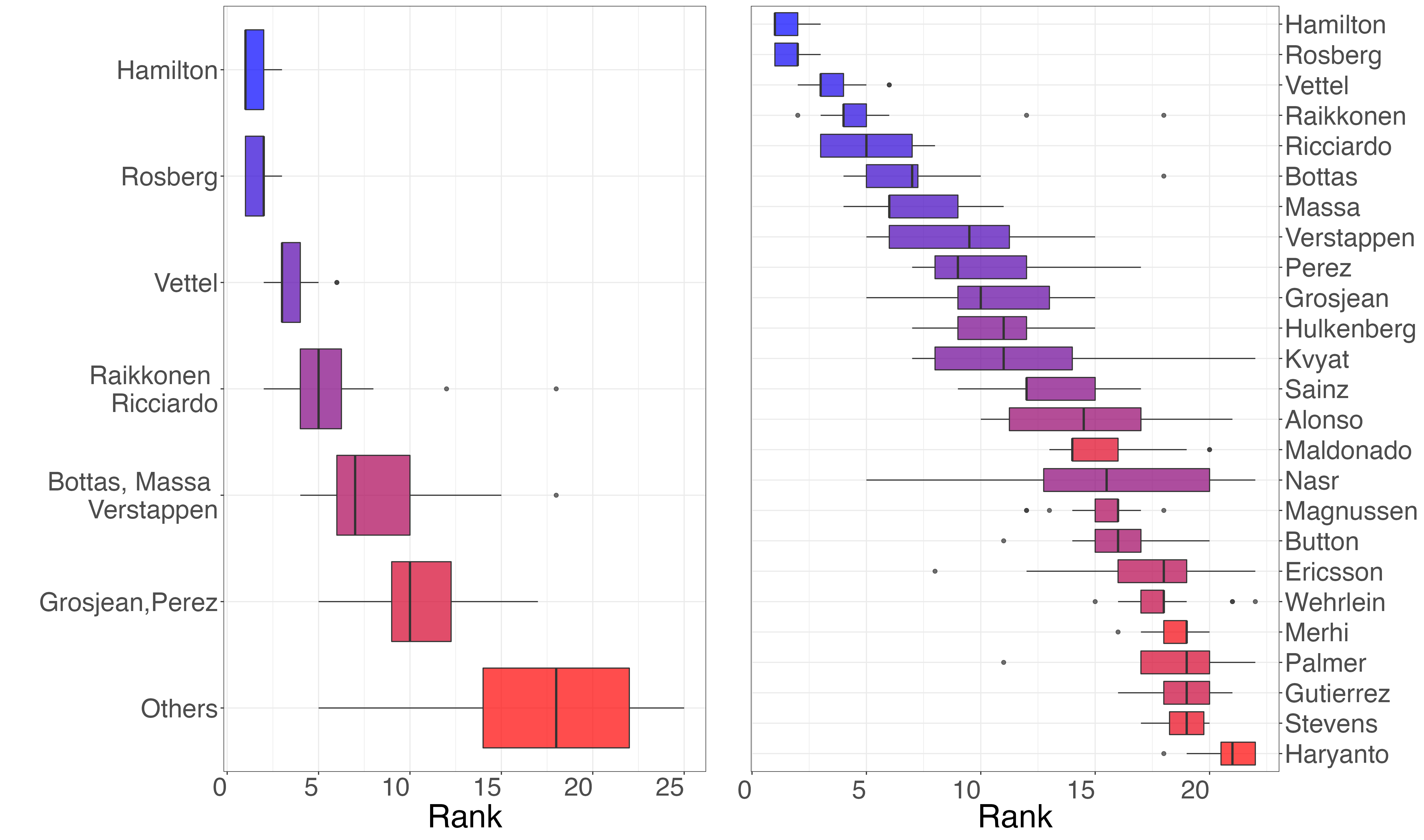}
    \caption{Distribution of clustered and unclustered ranks of drivers in Formula 1 seasons 2015 and 2016, corresponding to the CMM and MM, respectively.}
    \label{fig:boxes_F1}
\end{figure}

To select between models, the selection criterion (\ref{eq:info_criteria}) is computed per MCMC iteration with the current complete data and model parameters. Its maximum is $-1901.93$ for the CMM fit and $-2388.607$ in the MM. The value of $\mathcal{I}$ is higher for the CMM, indicating that the clustered model optimizes the penalised observed data probability. This claim is supported by Figure \ref{fig:boxes_F1} which displays the distribution of ranks by driver (MM) or driver subgroup (CMM). It is evident that aggregation suggested decreases substantially the uncertainty regarding how competitors are ordered.

\subsection{Tohoku sushi rankings}\label{sec:sushi}

The complete methodology developed throughout the paper is now used to analyse the Tohoku sushi rankings data introduced in section (\ref{sec:intro}). This is the collection of $q = 280$ rankings of $n=10$ sushi varieties provided by participants of the survey \cite{sushi} who lived in Tohoku until 15 years old. We pursue answers to the following questions. \textit{Is there a strict consensus between all sushi varieties in Tohoku?} and, if not, \textit{in what part of the ranking is there indifference, and between which sushis?}

First, Algorithm (\ref{alg:search}) is used to search for a CT applying one adjacent shifts. The search is done with $d_{og, h},d_{og, k}$ and $\mathcal{I}(\cdot)$ or $\mathcal{D}(\cdot, \underline{\pi})$. All combinations of distance and criteria take the initial $CT$ value obtained via the procedure of section \ref{sec:greedy_search}. The optimum $CT$s are shown in Table \ref{tab:tohoku_selection}.

\begin{table}[]
\centering
\begin{tabular}{@{}lcc@{}}
\toprule
$d_{go}$ & $\widehat{\mathcal{I}}(\cdot)$ & $\mathcal{D}(\cdot, \underline{\pi})$ \\ \midrule
Kendall  & (1,1,1,2,2,3)        & (1,2,2,5)                             \\
Hamming  & (1,2,2,2,3)          & (1,2,3,4)                             \\ \bottomrule
\end{tabular}
\caption{CMM structure selection for the Tohoku sushi rankings carried out with criteria $\mathcal{\widehat{I}}(\cdot)$ and $\mathcal{D}(\cdot, \underline{\pi})$. The likelihood-based information criterion is estimated with $M=100K$.}\label{tab:tohoku_selection}
\end{table}

\begin{figure}
    \centering
    \includegraphics[width = 0.6\linewidth]{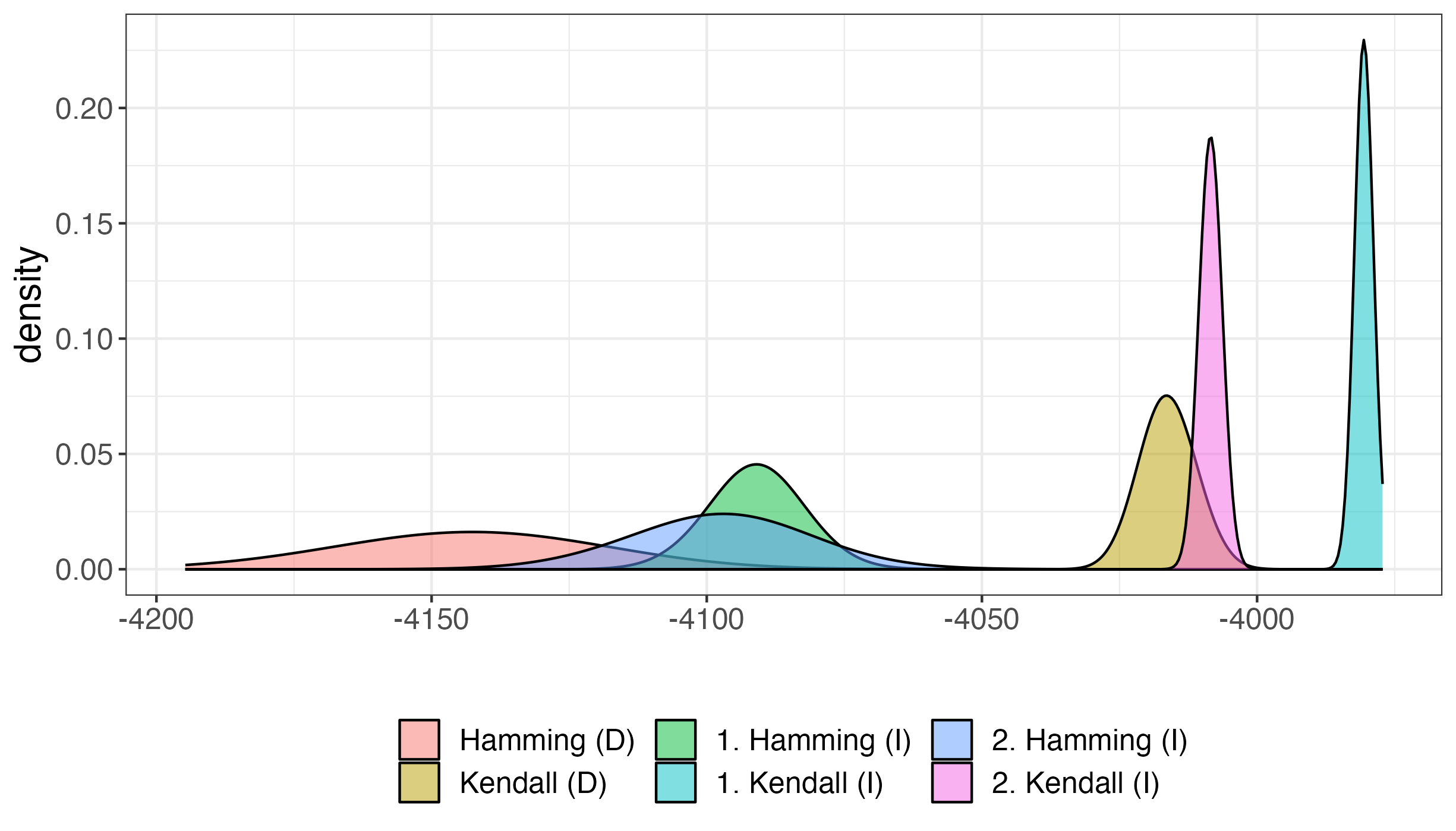}
    \caption{Distribution of $\widehat{\mathcal{I}}(\cdot)$ under candidate $CT$s estimated via Monte Carlo. The first and second-to-best models from the $CT$ search that uses $\widehat{\mathcal{I}}(\cdot)$ is considered ( \textit{1. Kendall (I), 2. Kendall (I), 1. Hamming (I), 2. Hamming (I)}) alongside those obtained by searching via the data-based criterion (\textit{Kendall (D), Hamming (D)}). }
    \label{fig:I_repetitions}
\end{figure}

As indicated by the simulation studies in section \ref{sec:greedy_search}, $\widehat{\mathcal{I}}(\cdot)$ more often identifies the true data-generating partition. Hence, it should be preferred when it is possible to obtain it with adequate precision. We evaluate its decisiveness by estimating the uncertainty of $\widehat{\mathcal{I}}(\cdot)$ under $M=100K$ via Monte Carlo replication. The criterion is recomputed 500 times for the models in Table \ref{tab:tohoku_selection} plus the second-to-best ones according to $\widehat{\mathcal{I}}(\cdot)$. Monte Carlo replication of each of these CTs with $M=100K$ yields the results used to produce Figure \ref{fig:I_repetitions}. This plot indicates that we can decisively select the model \textit{Kendall (I)} $\equiv (1,1,1,2,2,3)$. This $CT$ optimizes the likelihood-based criterion and has small enough variability to make the decision unambiguous.

Given that a model structure is selected, we fit its parameters using the Bayesian approach and the methodology of section \ref{sec:bayes}. We tackle the comparison between the clustered and unclustered models by fitting the latter (MM) with the $\texttt{BayesMallows}$ package. The CMM fit takes the proposals $\texttt{switch}(\cdot, 2)$ for $\boldsymbol{z}$, and log-Normal($\log \theta^t, \sigma$) with vanishing calibration of $\sigma$ for $\theta$. A priori, $\theta \sim \mbox{Gamma}(s =2, r=2)$ and $N$ is set to 200 for approximate exchange algorithm moves. The maximum a posteriori (MAP) parameters of the clustered and unclustered allocations are computed and displayed in Table \ref{tab:sushi_modes}. They differ in that the sushis $\{\texttt{sea urchin, salmon roe}\}$, $\{\texttt{sea eel, tuna roll, squid}\}$ and $\{\texttt{egg, cucumber roll}\}$ are indifferent within ranks $\{4,5\}$, $\{6,7,8\}$, $\{9,10\}$ according to the CMM, and follow the ordering of column 3 in the MM. 

For comparison with the MM, the information criterion (\ref{eq:info_criteria}) is recomputed for the CMM with the selected structure using the MAP parameter estimates. One hundred replications with $M=100K$ give an average $\widehat{\mathcal{I}}(CT)$ of $-3968.38$ with standard deviation $3.74$. This can be computed exactly for the Kendall-MM using the closed form of its normalisation constant. Its value is $-4341.95$ which indicates that the CMM is preferable. This can be affirmed decisively once the difference between the MM and the CMM $\mathcal{I}$ values is much larger than the CMM's $\hat{\mathcal{I}}$ standard deviation. Moreover, the decision is the same if we consider the unpenalised log-likelihood, allowing us to conclude that a tied consensus optimizes the probability of the observed data. 

\begin{table}[]
\centering
\begin{tabular}{@{}ccc@{}}
\toprule
\textbf{Ranks} & \textbf{CMM}                                                                  & \textbf{MM}                                                                    \\ \midrule
1              & \texttt{fatty tuna}                                          & \texttt{fatty tuna}                                           \\
2              & \texttt{tuna}                                                & \texttt{tuna}                                                 \\
3            & \texttt{shrimp}  & \texttt{shrimp}              \\

4,5 & \texttt{sea urchin} $\sim$ \texttt{salmon roe} & \texttt{sea urchin} $\succ$ \texttt{salmon roe} \\

6,7,8            & \texttt{sea eel} $\sim$ \texttt{tuna roll}  $\sim$ \texttt{squid}  &  \texttt{sea eel} $\succ$ \texttt{tuna roll}  $\succ$ \texttt{squid}  \\
9,10  &   \texttt{egg} $\sim$ \texttt{cucumber roll}    &  \texttt{egg} $\succ$ \texttt{cucumber roll} \\ \bottomrule
\end{tabular}
\caption{Maximum a posteriori estimates of the CMM $\boldsymbol{z}$ and MM $\pi_0$ fitted to Tohoku sushi rankings. In the CMM, $\boldsymbol{z}$ encodes an ordered partition where indifferent items are grouped in rows 4,5 and 6. Their strict order in $\pi_0$ is shown in the third column.}\label{tab:sushi_modes}
\end{table}



The alternative partitions are also compared in Table \ref{tab:correct_pref_props} using the data-based metric $\mathcal{D}(\cdot, \underline{\pi})$ by group. It displays the proportion of times that objects with lower cluster labels (or lower position in $\pi_0$) proceed those with higher labels in the data. Naturally, the three first columns are identical for the two models, since the first three CMM clusters are of single objects. Subsequently, there is an increase in the proportion of correct relations in the CMM, with an average of 0.68 for (4,5+), (5,6) which becomes 0.65 in the MM considering (4,5+), $\ldots$, (9,10).  

\begin{table}[]
\centering
\begin{tabular}{llllllllll}
\toprule
\multirow{2}{*}{CMM} & (1, 2+) & (2, 3+) & (3, 4+) & (4, 5+) & (5, 6)  &         &         &         &        \\  
                     & 0.74    & 0.67    & 0.68    & 0.64    & 0.73    &         &         &         &        \\ \hline
\multirow{2}{*}{MM}  & (1, 2+) & (2, 3+) & (3, 4+) & (4, 5+) & (5, 6+) & (6, 7+) & (7, 8+) & (8, 9+) & (9,10) \\
                     & 0.74    & 0.67    & 0.68    & 0.61    & 0.65    & 0.65    & 0.65    & 0.72    & 0.63   \\ \bottomrule 
\end{tabular}
\caption{Proportion of times that objects with lower cluster labels (or lower position in $\pi_0$) proceed those with higher labels in the sushi preferences data. This is computed between CMM clusters in the first row, between objects in the second row for the MM, and averaged by the number of comparisons. }\label{tab:correct_pref_props}
\end{table}

The Bayesian model fit estimates the posterior distribution of $\theta$, a parameter that quantifies the spread around some allocation $\boldsymbol{z}$. One useful way to look at this is with the MAP $\dot{\widehat{\boldsymbol{{z}}}}$ and rank-cluster probabilities. As shown in Equation (\ref{rank_type_probs}), the prevalence of clusters in rankings is a CMM expectation. Using percentage points of the $\theta$ posterior, we can construct confidence bands for $RC(i,l)$ and show uncertainty around it. The 2.5\% and 97.5\% quantiles of the distribution are $Q_{\theta}(2.5\%) =0.244$ and $Q_{\theta}(97.5\%)= 0.291$, so $P(0.244 < \theta < 0.291) = 0.95$. We can estimate 95\% bands for $RC(i,l)$ with the expected values $E_{f(\cdot|\dot{\widehat{\boldsymbol{{z}}}},Q_{\theta}(2.5\%))}[I\{ \widehat{\dot{z}}(\pi(i)) = l\} ]$, $E_{f(\cdot|\dot{\widehat{\boldsymbol{{z}}}},Q_{\theta}(97.5\%))}[I\{ \widehat{\dot{z}}(\pi(i)) = l\} ]$ estimated with Monte Carlo. Results obtained with 100K simulated realisations and some $i, l$ combinations are the following. The probability that \texttt{fatty tuna} ($l=1$) is the top item ($i=1$) is between 0.239 and 0.273. For \texttt{egg, cucumber roll}, this is (0.055,0.047) so either of these items has a very small probability of being the favorite. Any pair $i,l$ can be used in this analysis to obtain readily uncertainty quantification of group probabilities within ranks.

\section{Conclusions}

In this paper, the Clustered Mallows Model (CMM) was introduced as an extension of the well-known Mallows Model (MM). The CMM addresses the challenge of modelling rank collections that do not convey strict preferences among all items of interest. To do this, the CMM incorporates a population consensus parameter that can accommodate ties. Unlike the MM, which utilizes a modal permutation to fit $\underline{{\pi}}$, the CMM fits ordered clusters. Ordered clusters are achieved with the introduction of an item allocation vector, where we impose that within-group items are indifferent and that cluster labels hold an ordering interpretation. The CMM is motivated by a real data example of sushi rankings, where the relative preference between certain variants in the item set is unclear. By introducing the CMM, we provide a more comprehensive modeling approach for rank collections that encompass ambiguous preferences.

The CMM probability was defined in terms of a distance between ordered clusters, and a clustering table (CT). To determine suitable distance options, we referred to the extended metrics by \cite{critchlow}. These were formulated as CMM metrics and the model taking different distances were examined and compared. The resulting model has a non-analytical normalisation constant, which places the CMM posterior model in the class of doubly-intractable distributions. Accordingly, special attention was devoted to the problems of model selection and posterior sampling. The first was handled with an importance sampling estimator of the intractable term, which allowed for a model selection criterion to be evaluated. Posterior sampling was fully developed with Metropolis-Hastings and Approximate Exchange Algorithm (AEA, \citet{caimofriel2011}) moves that draw from the parameters' full-conditional distributions. With the application of the AEA, we were able to overcome the unavailability of CMM the normalisation term and fit the model's posterior distribution for $\underline{\boldsymbol{\pi}}$.

Bayesian inference in the presence of incomplete ranks was also addressed using data augmentation techniques. The CMM applicability to incomplete ranks was illustrated in an example that concerns results races in the Formula 1 competition. We were able to demonstrate that grouping drivers according to some scoring ranges allowed for a more robust representation of the data, which is preferred over the unclustered (MM) fit. We also outlined a greedy search of the CMM structure that relies on optimization of the penalised log-likelihood or a data-based goodness of fit metric.

A complete analysis of the motivating data was presented and the process of learning the CMM CT and model parameters was demonstrated in practice. The CMM and MM model fits were compared and we were able to show that the aggregation of some sushis in middle and lower-ranks can better describe the data-generating process.

Finally, some points of future research that deserve attention are the following. A CMM extension to the simultaneous clustering of assessors and items can be developed using a similar approach to \cite{vitelli2018}. This would consist of modelling clusters of rankers with specific CMM distributions. In turn, this would allow for fitting non-homogeneous populations with cluster(of assessors)-specific item partitions. Another point for future investigation is model selection, where sophisticated CT explorations could be stipulated with the TABU or genetic metaheuristics.

\section*{Acknowledgments}
This publication has emanated from research conducted with the financial support of Science Foundation Ireland under Grant number 18/CRT/6049.  
The Insight Centre for Data Analytics is supported by Science Foundation Ireland under Grant Number 12/RC/2289$\_$P2. The authors report there are no competing interests to declare.


\newpage





\bibliographystyle{apalike}
\bibliography{bib}

\end{document}